\documentclass[12pt]{article}
\begin{document}

\begin{center}
\bf{Entanglement structure of adjoint representation of unitary
group and tomography of quantum states}
\end{center}


E-mail:

manko@sci.lebedev.ru

marmo@na.infn.it

sudarshan@physics.utexas.edu

zaccaria@na.infn.it

Submitted in English 1 August 2003

\section*{Abstract}

The density matrix of composite spin system is discussed in
relation to the adjoint representation of unitary group $U(n)$.
The entanglement structure is introduced as an additional
ingredient to the description of the linear space carrying the
adjoint representation. Positive maps of density operator are
related to random matrices. The tomographic probability
description of quantum states is used to formulate the problem of
separability and entanglement as the condition for joint
probability distribution of several random variables represented
as the convex sum of products of probabilities of random variables
describing the subsystems. The property is discussed as a possible
criterion for separability or entanglement. The convenient
criterion of positivity of finite and infinite matrix is obtained.
The $U(n)$-tomogram of a multiparticle spin state is introduced.
The entanglement measure is considered in terms of this tomogram.

KEY WORDS: unitary group, entanglement, adjoint representation,
tomogram, operator symbol, random matrix.

\section{Introductiion}

The notion of entanglement \cite{Schroedinger} is related to the
quantum composition principle of the states of subsystems for a
given multipartite system.
For pure states, the notion of entanglement and separability can
be given as follows.

If the wave function of a state of a bipartite system is
represented as the product of two wave functions depending on
coordinates of the subsystems, the state is simply separable;
correspondingly, in other cases, the state is entangled. An
intrinsic approach to the entanglement measure was suggested in
\cite{MMSZ-J.Phys.A}. The measure was introduced as the distance
between the system density matrix and the tensor product of the
subsystem states. There are several other different
characteristics and measures of entanglement considered by several
authors [3--9].
 Each of the entanglement measures
describes a degree of correlations between the subsystems'
properties. The notion of entanglement is not an absolute notion
for a given system but depends on the decomposition into
subsystems. The same quantum state can be considered as entangled,
if one kind of division of the system into subsystems is given, or
as completely disentangled, if another decomposition of the system
into subsystems is considered.

For instance, the state of two continuous quadratures can be
entangled in Cartesian coordinates and disentangled in polar
coordinates. Coordinates are considered as measurable observables
labeling the subsystems of the given system. The choice of
different subsystems mathematically implies the existence of two
different sets of the subsystems' characteristics (we focus on
bipartite case). We may consider the Hilbert space of states
$H(1,2)$ or $H(1^{\prime },2^{\prime })$. The  Hilbert space for
the total system is, of course, the same but the index $(1,2)$
means that there are two sets of operators $P_{1}$ and $P_{2}$,
which select subsystem states 1  and 2. The index $(1^{\prime
},2^{\prime })$ means that there are other two sets of operators
$P_{1}^{\prime }$ and  $P_{2}^{\prime }$, which select subsystem
states $1^{\prime }$ and $2^{\prime }.$  The operators $P_{1,2}$
and $P_{1^{\prime },2^{\prime }}^{\prime }$ have specific
properties. They are represented as tensor products of operators
acting in the space of states of the subsystem 1 (or 2) and unit
operators acting in the subsystem 2 (or 1). In other words, we
consider the space $H$, which can be treated as tensor product of
spaces  $H(1)$ and $H(2)$ or $H(1')$ and $H(2')$. In the
subsystems $1$ and $2$, there are basis vectors $\mid
n_{1}\rangle$ and $ \mid m_{2}\rangle$, as well as in the
subsystems $1^{\prime }$ and  $2^{\prime }$ there are basis
vectors $\mid n_{1}^{\prime }\rangle$ and  $\mid m_{2}^{\prime
}\rangle.$ The vectors  $\mid n_{1}\rangle \mid m_{2}\rangle$ and
the vectors $\mid n_{1}^{\prime }\rangle\mid
m_{2}^{\prime}\rangle$ form the sets of basis vectors in the
composite Hilbert space, respectively. These two sets are related
by means of unitary transformation. An example of such a composite
system is a bipartite spin system.

If one has spin-$j_{1}$ [the space $H(1)$] and spin-$j_{2}$ [the
space $H(2)$] systems, the combined system can be treated as
having basis $\mid j_{1}m_{1}\rangle\mid j_{2}m_{2}\rangle.$

Another basis in the composite-system-state space can be
considered in the form  $\mid j_{{}}m_{{}}\rangle$, where $j$ is
one of the numbers $|j_{1}-j_{2}|,$  $|j_{1}-j_{2}|+1,\ldots,$
$j_{1}+j_{2}$ and  $m=m_{1}+m_{2}$. The basis $\mid jm\rangle$ is
related to the basis $\mid j_{1}m_{1}\rangle\mid
j_{2}m_{2}\rangle$ by means of unitary transform given by
Clebsch--Gordon coefficients $C$ $(j_{1}m_{1} j_{2}m_{2}|jm)$.
From the viewpoint of given definition, the states $\mid
j_{{}}m_{{}}\rangle$ are entangled states. For example, if
$j_{1}=j_{2}={1}/{2}$, there are entangled spin states of the
composite system, which nowadays are called Bell states $$\mid
\Phi^{\pm }\rangle=\frac{1}{\sqrt{2}}\left(\left|\frac{1}{2}
\frac{1}{2}\right.\Big\rangle_{1}\left|\frac{1}{2}\frac{1}{2}
\right.\Big\rangle_{2}\pm\left|\frac{1}{2}\frac{-1}{2}\right.
\Big\rangle_{1}\left|\frac{1}{2}\frac{-1}{2}\right.\Big\rangle_{2}
\right),$$
$$|\Psi^{\pm}\rangle=-\frac{1}{\sqrt{2}}\left(\left|\frac{1}{2}\frac{1}{2}
\Big\rangle_{1}\right.\left|\frac{1}{2}\frac{-1}{2}\Big\rangle_{2}
\right.\pm\left|\frac{1}{2}\frac{-1}{2}\Big\rangle_{1}\right.\left|
\frac{1}{2}\frac{1}{2}\Big\rangle_{2}\right.\right).$$ These
states are maximally entangled states. In terms of spin, the
states $\mid\Phi ^{\pm }\rangle$ are the superpositions of $j=1,$
$ m=\pm 1$ states and the states $\mid\Psi ^{\pm }\rangle$ are the
superpositions of $j=0, 1$, $m=0$ states.

The spin states can be described by means of the tomographic
map~[10--12].
 For bipartite spin systems, the states were described  by the
tomographic probabilities in \cite{Andreev,Andreev-Olga}. Some
properties of the tomographic spin description were studied in
\cite{Marmo}.  In the tomographic approach, the problems of the
quantum state entanglement can be cast into the form of some
relations among the probability distribution functions. On the
other hand, to have a clear picture of entanglement, one needs
mathematical formulation of properties of the density matrix of
the composite system, a description of the linear space of the
composite system states. Since the density matrix is hermitian,
the space of states is a subset of linear space of adjoint
representation of the group $U(n^{2})$, where $n=2j+1$ is the
dimension of the spin states of two spinning particles. Thus one
needs to characterize the connection of the entanglement phenomena
with the structures in the space of adjoint representation of the
$U(n^{2})$ group.

The aim of this paper is to connect entanglement problems with the
properties of tomographic probability distributions and discuss
the properties of the convex set of positive states for composite
system with taking into account the subsystem structures. We used
Hilbert--Schmidt distance to calculate the measure of entanglement
as the distance between a given state and the tensor product of
the partial traces of the density matrix of the given state. In
\cite{DeCastro} another measure of entanglement as a
characteristic of subsystem correlations was introduced. This
measure is determined via covariance matrix of some observables.
Review of different approaches to the entanglement notion and
entanglement measures is given in \cite{J.Math.Phys:SI-Sept.2002},
where the approach to describe entanglement and separability of
composite systems is based on entropy methods.

Due to variety of approaches to the entanglement problem, one
needs to understand better what in reality this word
`entanglement' describes. Is it a synonym of the word
`correlation' between two subsystems or it has to capture some
specific correlations attributed completely and only to the
quantum domain?

The paper is organized as follows.

In Sec. 2 we study the subsystem structure of a given linear
space. In Sec. 3 we consider the relation of the group $U(n^{2})$
to the set of density matrices. In Sec. 4 we discuss positive
maps. In Sec. 5 we investigate local transforms. In Sec. 6 we
treat the probability distributions as vectors. In. Sec. 7 we
prove the invariance of the intrinsic entanglement measure. In
Sec. 8 we define the separable states. In Sec. 9 generic symbols
of operators are presented. In Sec. 10 an example of Weyl symbols
is considered and in Sec. 11 an example of quadrature tomogram is
done. In Sec. 12 symbols of density operators for multipartite
system are discussed. Spin tomography is reviewed in Sec. 13. Two
qubits are considered in the tomographic representation in Sec.
14. The relation of dynamical map to purification procedure is
described in Sec. 15. Some properties of quadratic forms are
reviewed in Sec. 16. The tomogram for the group $U(N)$ is
introduced in Sec. 17. Conclusions and results are listed in Sec.
18.

\section{Linear Space of a Composite System, Its Structure, and
Its Convex Subset of Positive States}

In this section, we review the meaning and notion of composite
system in terms of additional structures on the linear space of
state for the composite system.

\subsection{States and Observables}

In quantum mechanics, there are two principal ingredients, which
are associated with linear operators acting in a Hilbert space.
The first ingredient is related to the concept of quantum state
and the second one, to the concept of observable. The state is
associated to Hermitian nonnegative, trace-class, linear operator.
The observables are associated to Hermitian operators. Though the
both states and observables are identified with the Hermitian
operators, there is an essential difference between these two
objects. The observables have additional product structure. Thus
we consider product of two linear Hermitian operators
corresponding to the observables. First measuring an observable
and (after measuring the first one) measuring another observable
just correspond to the product of two operators.

For the states, the notion of product is redundant. The product of
two states is not a state. For states, one keeps only the linear
structure of vector space. For finite $n$-dimensional system, the
Hermitian states and the Hermitian observables live in Lie algebra
of the unitary group $U(n)$. But the states correspond to
nonnegative Hermitian operator. The observables can be associated
with both types of the operators including nonnegative and
nonpositive ones. Space of states is linear space which, in
principle, is not equipped by a product structure. Due to this, if
one considers transformations in linear space of states, one does
not need to preserve any product structure. In the set of
observables, one needs to care what is happening with product of
operators provided some transformations are applied.

\subsection{Vectors}

Let us first introduce some extra constructions of the map of a
matrix onto a vector. Given a rectangular matrix $M$ with elements
$M_{id}$, where $i=1,2,\ldots,n$  and  $d=1,2,\ldots,m$. Then one
can consider the matrix as a vector $\mathcal{\vec{M}}$ with
$N=nm$ components constructed by the following rule:
\begin{equation}
\label{eq.1}
 \mathcal{M}_{1}=M_{11}, \quad\mathcal{M}_{2}=M_{12},
\quad\mathcal{M}_{m}=M_{1m},\quad\mathcal{M}_{m+1}=M_{21},\ldots
\mathcal{M}_{N}=M_{nm}.
\end{equation}
Thus we construct the map $M\rightarrow
\mathcal{\vec{M}=}\hat{t}_{\mathcal{\vec{M}}M}M.$

We have introduced the linear operator
$\hat{t}_{\mathcal{\vec{M}}M}$  which maps the matrix $M$ on a
vector $\mathcal{\vec{M}}$. Now we introduce the inverse operator
$\hat{p}_{\mathcal{\vec{M}}M}$ which maps a given vector column in
the space with dimension $N=mn$  onto the rectangular matrix. This
means that given a vector
$\mathcal{\vec{M}=M}_{1},\ldots,\mathcal{M}_{N}$, we use the rule
of relabeling the components of the vector introducing two indices
$i=1,\ldots,n$ and $d=1,\ldots,m.$  The relabeling is accomplished
according to (\ref{eq.1}).  Then we collect the relabeled
components into matrix table. Thus we get the map
\begin{equation}\label{eq.2}
\hat{p}_{\mathcal{\vec{M}}M}\mathcal{\vec{M}}=M.
\end{equation}
One can see that their composition
\begin{equation}\label{eq.3}
\hat{t}_{\mathcal{\vec{M}}M}\hat{p}_{\mathcal{\vec{M}}M}
\mathcal{\vec{M}=}1\cdot\mathcal{\vec{M}}
\end{equation}
acts on the vector as unit operator in the linear space of
vectors.

Given a $n$$\times$$n$ matrix the map suggested can also be
extended. The matrix can be treated as $n^{2}$-dimensional vector
and, vice versa, the vector of dimension $n^{2}$ can be mapped by
this procedure onto the $n$$\times$$n$ matrix.

Let us consider a linear operator acting on the vector $\vec{\cal
M}$ and related to a linear transform of the matrix $M$. First, we
study the correspondence of the linear transform of the form
\begin{equation}\label{L1}
M\rightarrow gM=M_g^l
\end{equation}
to the transform of the vector
\begin{equation}\label{L2}
\vec{\cal M}\rightarrow \vec{\cal M}_g^l={\cal L}_g^l\vec {\cal
M}.
\end{equation}
One can show that the $n^2$$\times$$n^2$ matrix ${\cal L}_g^l$ is
determined by the tensor product of the $n$$\times$$n$ matrix $g$
and $n$$\times$$n$ unit matrix, i.e.,
\begin{equation}\label{L3}
{\cal L}_g^l=g\otimes 1.
\end{equation}
Analogously, the linear transform of the matrix $ M$ of the form
\begin{equation}\label{L4}
M\rightarrow Mg=M_g^r
\end{equation}
induces the linear transform of the vector $\vec{\cal M}$ of the
form
\begin{equation}\label{L5}
\vec{\cal M}\rightarrow \vec{\cal M}_g^r=\hat t_{\vec {\cal
M}{\cal M}}M_g^r={\cal L}_g^r\vec {\cal M},
\end{equation}
where the $n^2$$\times$$n^2$ matrix ${\cal L}_g^r$ reads
\begin{equation}\label{L6}
{\cal L}_g^r=1\otimes g^{\mbox{tr}}.
\end{equation}
Similarity transformation of the matrix $M$ of the form
\begin{equation}\label{L7}
M\rightarrow gMg^{-1}
\end{equation}
induces the corresponding linear transform of the vector
$\vec{\cal M}$ of the form
\begin{equation}\label{L8}
\vec{\cal M}\rightarrow \vec{\cal M}_s={\cal L}_g^s\vec {\cal M},
\end{equation}
where the $n^2$$\times$$n^2$ matrix ${\cal L}_g^s$ reads
\begin{equation}\label{L9}
{\cal L}_g^s=g\otimes (g^{-1})^{\mbox{tr}}.
\end{equation}
One can ask how to determine the inverse map of vector
$\vec{\cal M}$ onto matrix $M$, i.e., how to define the operator
$\hat p_{\vec{\cal M}M}$. In fact, the reconstruction can be
defined by means of star-product of vectors $\vec{\cal M}$ in a
linear space. One can define the associative product of two
$N$-vectors $\vec{\cal M}_1$ and $\vec{\cal M}_2$ using the rule
\begin{equation}\label{L10}
\vec{\cal M}=\vec{\cal M}_1\star\vec{\cal M}_2,
\end{equation}
where
\begin{equation}\label{L11}
\vec{\cal M}_k=\sum_{l,s=1}^NK_{ls}^k(\vec{\cal M}_1)_l(\vec{\cal
M}_2)_s.
\end{equation}
If one applies a linear transform to the vectors $\vec{\cal
M}_1$, $\vec{\cal M}_2$, $\vec{\cal
 M}$ of the form
$$ \vec{\cal M}_1\rightarrow\vec{\cal M}^\prime_1= {\cal
L}\vec{\cal M}_1,\qquad \vec{\cal M}_2\rightarrow\vec{\cal
M}^\prime_2= {\cal L}\vec{\cal M}_2,\qquad \vec{\cal
M}\rightarrow\vec{\cal M}^\prime= {\cal L}\vec{\cal M}, $$ the
invariance of the star-product kernel yields $$ \vec{\cal
M}_1^\prime \star\vec{\cal M}^\prime_2= \vec{\cal M}^\prime,\qquad
\mbox{if}\qquad {\cal L}=G\otimes G^{-1\mbox{tr}},\quad G\in
GL(n).$$ The kernel $K_{ls}^k$ (structure constants) which
determines the associative star-product satisfies the quadratic
equation. Thus if one wants to make the correspondence of the
vector star-product to the standard matrix product (row by
column), the matrix $M$ must be constructed appropriately. For
example, if the vector star-product is commutative, the matrix $M$
corresponding to the $N$-vector $\vec{\cal M}$ can be chosen as
diagonal $N$$\times$$N$ matrix. This consideration shows that the
map of matrices on the vectors provides star-product of the
vectors (defines the structure constants or the kernel of
star-product) and, conversely, if one has the vectors, the map of
the vectors onto the matrices with the standard multiplication
rule is determined by the structure constants (or by the kernel of
the vector star-product).

The constructed map of matrices on the vectors gives a possibility
to enlarge the dimensionality of the group acting in the linear
space of matrices in comparison with the standard one. Thus, given
a $n$$\times$$n$ matrix $M$ the left action, the right action, and
similarity transformation of the matrix are related to the complex
group $GL(n)$. On the other hand, the linear transformations in
the linear space of $n^2$-vectors $\vec{\cal M}$ obtained by using
the introduced map are determined by the matrices belonging to the
group $GL(n^2)$. There are transformations on the vectors which
cannot be \underline{simply} represented on matrices. If
$M\to\Phi(M)$ is a linear homogeneous function of the matrix $M$,
we may represent it by $$\Phi_{ab}=B_{aa',\,bb'}M_{a'b'}.$$ Under
rather clear conditions, $B_{aa',bb'}$ can be expressed in terms
of its nonnormalized left and right eigenvectors:
$$B_{aa',bb'}=\sum_\nu x_{aa'}(\nu)y^\dagger_{bb'}(\nu), $$ which
corresponds to $$
\Phi(M)=xMy^\dagger=\sum_{\nu=1}^{n^2}x(\nu)My^\dagger(\nu).$$

There are possible linear transforms on the matrices and
corresponding linear transforms on the induced vector space which
belong not to a group but to an algebra of matrices. One can
describe the map of $n$$\times$$n$ matrices $M$ (source space)
onto vectors $\vec{\cal M}$ (target space) using specific basis in
the space of the matrices. The basis is given by the matrices
$E_{jk}~(j,k=1,2,\ldots,n)$ with all matrix elements equal to zero
except the element in $j$th row and $k$th column which is equal to
unity. One has the obvious property
\begin{equation}\label{L12}
M_{jk}=\mbox{Tr}\left(ME_{jk}\right).
\end{equation}
In our procedure, the basis matrix $E_{jk}$ is mapped onto the
basis column-vector $\vec{\cal E}_{jk}$, which has all components
equal to zero except the unity component related to the position
in the matrix determined by the numbers $j$ and $k$. Then one has
\begin{equation}\label{L13}
\vec{\cal M}=\sum_{j,k=1}^n\mbox{Tr}\left(ME_{jk}\right)\vec{\cal
E}_{jk}.
\end{equation}
For example, for similarity transformation of the finite matrix
$M$, one has
\begin{equation}\label{L14}
\vec{\cal M}_g^s=
\sum_{j,k=1}^N\mbox{Tr}\left(gMg^{-1}E_{jk}\right)\vec{\cal
E}_{jk}.
\end{equation}

Now we will define the notion of `composite' vector which
corresponds to dividing a quantum system into subsystems.

We will use the following terminology.

In general, the given linear space of dimensionality $N=mn$ has a
structure of bipartite system, if the space is equipped with the
operator $\hat{p}_{\mathcal{\vec{M}}M}$ and the matrix $M$
(obtained by means of the map) has matrix elements in factorizable
form
\begin{equation}\label{eq.4}
M_{id}\rightarrow x_{i}y_{d}.
\end{equation}
This $M=x\otimes y$ corresponds to the special case of
nonentangled states. Otherwise, one needs
$$M=\sum_{\nu}x(\nu)\otimes y(\nu).$$ In fact, to consider in
detail the entanglement phenomenon, in the bipartite system of
spin, one has to introduce a hierarchy of three linear spaces. The
first space of pure spin states is two-dimensional linear space of
complex vectors
\begin{equation}\label{new1a}
\mid \vec x\rangle=\left(
\begin{array}{c}
x_1 \\x_2
\end{array}
\right).  \end{equation} In this space, the scalar product is
defined as follows:
\begin{equation}\label{new2a}
\langle \vec x\mid\vec y\rangle=x_1^*y_1+x_2^*y_2.
\end{equation}
So it is two-dimensional Hilbert space. We do not equip this space
with a vector star-product structure. In the primary linear space,
one introduces linear operators $\hat M$ which are described by
2$\times$2 matrices $M$. Due to the map discussed in the previous
section, the matrices are represented by 4-vectors $\vec{\cal M}$
belonging to the second complex 4-dimensional space. Star-product
of the vectors $\vec{\cal M}$ determined by the kernel ${\cal
K}_{ls}^k$ is defined in such a manner in order to correspond to
the standard rule of multiplication of the matrices.

In addition to the star-product structure, we introduce the scalar
product of the vectors $\vec{\cal M}_1$ and $\vec{\cal M}_2$, in
view of the definition
\begin{equation}\label{new3a}
\langle \vec {\cal M}_1\mid\vec {\cal M}_2\rangle
=\mbox{Tr}\,(M_1^\dagger M_2),
\end{equation}
which is the trace formula for scalar product of matrices.

This means introducing the metric $g^{\alpha\beta}$ in the
standard notation for scalar product
\begin{equation}\label{new4a}
\langle \vec {\cal M}_1\mid\vec {\cal M}_2\rangle
=\sum_{\alpha,\beta=1}^4(M_1)^*_\alpha g^{\alpha\beta}(
M_2)_\beta,
\end{equation}
where the matrix $g^{\alpha\beta}$ is of the form
\begin{equation}\label{new5a}
g^{\alpha\beta}=\left(
\begin{array}{clcr}
1&0&0&0 \\
0&0&1&0 \\
0&1&0&0 \\
0&0&0&1
\end{array}
\right),\qquad g^{\alpha j}g^{j\beta}=\delta^{\alpha\beta}.
  \end{equation}
The scalar product is invariant under action of the group of
nonsingular 4$\times$4 matrices $\ell$, which satisfy the
condition
\begin{equation}\label{L22}
 \ell^{-1}=g\ell^\dagger g.
 \end{equation}
The product of matrices $\ell$ satisfies the same condition
since $g^2=1.$

Thus, the space of operators $\hat M$ in
primary two-dimensional space of spin states is mapped onto linear
space which is equipped with a scalar product (metric Hilbert space
structure) and associative star-product (kernel satisfying
quadratic associativity equation). In the linear space of the
4-vectors $\vec {\cal M}$, we introduce linear operators
(superoperators), which can be associated
with algebra of 4$\times$4 complex matrices.

\section{Density Operators and Positive Maps}

In this section, we focus on density matrices. This means that our
matrix $M$ is considered as a density matrix $\rho$ which
describes a quantum state. We consider here the action of the
unitary transformation $U(n)$ of the density matrices and
corresponding transformations on the vector space. If one has
structure of bipartite system, we also consider the action of
local gauge transformation both in the `source space' of density
matrices and in the `target space' of the corresponding vectors.

The $n$$\times$$n$ density matrix $\rho $ has matrix elements
\begin{equation}\label{eq.5}
\rho _{ik}=\rho _{ki}^{\dagger }, \qquad\mbox{Tr}\,\rho _{{}}=1,
\qquad\langle\psi |\rho |\psi\rangle \geq 0.
\end{equation}
Since the density matrix is hermitian, it can be always identified
as an element of the convex subset of the linear space associated
 with the Lie algebra
of $U(n)$ group, on which the group $U(n)$ acts with the adjoint
representation
\begin{equation}\label{eq.6}
\rho \rightarrow \rho _{U}=U\rho U_{{}}^{\dagger }.\end{equation}
The system is said to be bipartite, if the space of representation
is equipped with an additional structure. It means that for
$$n^{2}=n_{1}^{{}}\cdot n_{2}^{{}},\qquad n_{1}=n_{2}=n$$ one can
make first the map of $n$$\times$$n$ matrix $\rho _{{}}$ onto
$n^{2}$-dimensional vector $\vec{\rho}$ according to the previous
procedure, i.e., one equips the space by an operator
$\hat{t}_{\vec{\rho}\rho }$. Given this vector one makes a
relabeling of the vector $\vec{\rho}$ components according to the
rule
\begin{equation}\label{eq.7}
\vec{\rho}\rightarrow \rho _{id,ke}, \quad
i,k=1,2,\ldots,n_{1},\quad d,e=1,2,\ldots,n_{2}, \end{equation}
i.e., obtaining the quadratic matrix
\begin{equation}\label{eq.8}
\rho _{q}=\hat{p}_{\rho _{q}\vec{\rho}}\vec{\rho}. \end{equation}
The unitary transform (\ref{eq.6}) of the density matrix induces
the linear transform of the vector $\vec\rho$ of the form
\begin{equation}\label{new1}
\vec\rho\rightarrow \vec\rho _U=(U\otimes U^*)\vec\rho.
\end{equation}
There exist linear transforms (called positive maps) of the
density matrix, which preserve its trace, hermicity, and
positivity. It is the transform introduced in \cite{Sud61}
\begin{equation}\label{new2}
\rho_0\rightarrow \rho _s=\sum_kp_kU_k\rho_0U_k^\dagger,
\quad\sum_kp_k=1,
\end{equation}
where $U_k$ are unitary matrices and $p_k$ are positive numbers.

If the initial density matrix is diagonal, i.e., it belongs to
Cartan subalgebra of Lie algebra of the unitary group, the
diagonal elements of the obtained matrix give smoother probability
distribution than the initial one. There exists the generic
transform (see \cite{Sud61,geom})
\begin{equation}\label{new3}
\rho_0\rightarrow \rho=\sum_k V_k\rho_0V_k^\dagger,
\quad\sum_kV_k^\dagger V_k=1.
\end{equation}
For large number of terms in the sum, the above map gives the most
stochastic density matrix $$\rho_0\rightarrow \rho_s=(n)^{-1}1.$$
The transform (\ref{new2}) is the partial case of the transform
(\ref{new3}). We discuss the transforms separately since they are
used in the literature in the presented form.

One can see that the constructed map of density matrices onto
vectors provides the corresponding transforms of the vectors,
i.e.,
\begin{equation}\label{new4}
\vec\rho_0\rightarrow \vec\rho_s=\sum_kp_k(U_k\otimes
U_k^*)\vec\rho_0
\end{equation}
and
\begin{equation}\label{new5}
\vec\rho_0\rightarrow \vec\rho=\sum_k(V_k\otimes V_k^*)\vec\rho_0.
\end{equation}
It is obvious that the linear transforms of the vectors, which
preserve their properties to correspond to the density matrix, are
essentially larger than the standard unitary transform of the
density matrix.

Formulas (\ref{new4}) and (\ref{new5}) mean that the positive map
superoperators acting on the density matrix in the vector
representation are described by $n^2$$\times$$n^2$ matrices
\begin{equation}\label{new6}
{\cal L}_s=\sum_kp_k(U_k\otimes U_k^*)
\end{equation}
and
\begin{equation}\label{new7}
{\cal L}=\sum_kV_k\otimes V_k^*,
\end{equation}
respectively.

Positive map is called `noncompletely positive' if $${\cal
L}=\sum_kV_k\otimes V_k^*-\sum_sv_s\otimes v_s^*,\qquad
\sum_kV_k^\dagger V_k-\sum_sv_s^\dagger v_s=1.$$ This map is
related to nonphysical evolution of a subsystem.

\section{Positive Map and Random Matrices}

Formula (\ref{new6}) can be considered in the context of random
matrix representation. In fact, the matrix ${\cal L}_s$ can be
interpreted as the weighted mean value of the random matrix
$U_k\otimes U_k^*$. The dependence of matrix elements and positive
numbers $p_k$ on index $k$ means that we have a probability
distribution function $p_k$ and averaging of the random matrix
$U_k\otimes U_k^*$ by means of the distribution function. So the
matrix ${\cal L}_s$ reads
\begin{equation}\label{R1}
{\cal L}_s=\langle U\otimes U^*\rangle.
\end{equation}
Let us consider example of 2$\times$2 unitary matrix.
We can consider the matrix
 of $SU(2)$ group of the form
\begin{equation}\label{R2}
u=\left(
\begin{array}{cl}
\alpha&\beta\\
-\beta^*&\alpha^*
\end{array}
\right),\qquad |\alpha|^2+|\beta|^2=1.  \end{equation}
The 4$\times$4 matrix ${\cal L}_s$ takes the form
\begin{equation}\label{R3}
{\cal L}_s=\left(
\begin{array}{clcr}
\ell&m&m^*&1-\ell\\
-n&s&-q&n\\
-n^*&-q^*&s^*&n^*\\
1-\ell&-m&-m^*&\ell
\end{array}
\right).  \end{equation}
The matrix elements of the matrix ${\cal L}_s$ are the means
\begin{eqnarray}
m&=&\langle \alpha\beta^*\rangle,\nonumber\\ \ell&=&\langle
\alpha\alpha^*\rangle,\nonumber\\ n&=&\langle
\alpha\beta\rangle,\label{R4}\\ s&=&\langle
\alpha^2\rangle,\nonumber\\ q&=&\langle \beta^2\rangle.\nonumber
\end{eqnarray}
Moduli of these matrix elements are smaller than unity.

Determinant of the matrix ${\cal L}_s$ reads
\begin{equation}\label{R5}
\mbox{det}\,{\cal L}_s=(1-2\ell)\Big(|q|^2-|s|^2\Big)+4\,\mbox{Re}
\,\Big[q^*m^*n+mns^*\Big].
\end{equation}
If one represents the matrix ${\cal L}_s$ in block form
\begin{equation}\label{R6}
{\cal L}_s =\left(
\begin{array}{cl}
A&B\\
C&D\end{array}
\right),  \end{equation}
then
\begin{equation}\label{R7}
A =\left(
\begin{array}{cl}
\ell&m\\
-n&s\end{array}
\right),  \qquad
B =\left(
\begin{array}{cl}
m^*&1-\ell\\ -q&n\end{array} \right),  \end{equation} and
\begin{equation}\label{R8}
D=\sigma_2A^*\sigma_2,\qquad
C=-\sigma_2B^*\sigma_2,\end{equation}
where $\sigma_2$ is Pauli matrix.

One can check that the product of two different matrices
${\cal L}_s$ can be cast in the same form. This means that the matrices
${\cal L}_s$ form the 9-parameter compact semigroup. For example,
in the case $\ell=1/2$ and $m=0$, one has the matrices
\begin{equation}\label{R10}
A =\left(
\begin{array}{cl}
1/2&0\\
-n&s\end{array}
\right),  \qquad
B =\left(
\begin{array}{cl}
0&1/2\\
-q&n\end{array}
\right).  \end{equation}
Determinant of the matrix ${\cal L}_s$ in this case is equal to zero.
All the matrices ${\cal L}_s$ have the eigenvector
\begin{equation}\label{R11}
\vec\rho_0=\left(
\begin{array}{c}
1/2\\0\\0\\1/2\end{array}
\right),  \end{equation}
i.e.,
\begin{equation}\label{R12}
{\cal L}_s\vec\rho_0=\vec\rho_0.\end{equation}
This eigenvector corresponds to the density matrix
\begin{equation}\label{R13}
\rho_1=\left(
\begin{array}{cl}
1/2&0\\
0&1/2\end{array}
\right),  \end{equation}
which is obviously invariant of the positive map.

For random matrix, one has correlations of the random matrix elements,
e.g., $\langle \alpha\alpha^*\rangle\neq\langle
\alpha\rangle\langle\alpha^*\rangle.$

The matrix ${\cal L}_p$
\begin{equation}\label{R14}
{\cal L}_p=\left(
\begin{array}{clcr}
1&0&0&0\\
0&0&1&0\\
0&1&0&0\\
0&0&0&1
\end{array}
\right)  \end{equation}
maps the vector
\begin{equation}\label{R15}
\rho_{\rm in}=\left(
\begin{array}{c}
\rho_{11}\\\rho_{12}\\\rho_{21}\\\rho_{22}\end{array}
\right)  \end{equation}
onto the vector
\begin{equation}\label{R16}
\vec\rho_{\rm t}=\left(
\begin{array}{c}
\rho_{11}\\\rho_{21}\\\rho_{12}\\\rho_{22}\end{array} \right).
\end{equation} This means that the positive map (\ref{R14})
connects the positive density matrix with its transposed. This map
can be presented as the connection of the matrix $\rho$ with its
transposed of the form
$$\rho\to\rho^T=\rho^*=\frac{1}{2}\,\Big(\rho+\sigma_1\rho\sigma_1
-\sigma_2\rho\sigma_2+\sigma_3\rho\sigma_3\Big).$$ There is no
unitary transform connecting these matrices.

This noncompletely positive map in $N$-dimensional case is given
by generalized formula $$\rho\rightarrow\rho_s=-\varepsilon
\rho+\frac{1+\varepsilon}{N}\,1_N,\qquad \varepsilon>0.$$ The
standard unitary transform can be interpreted as average random
transform with probability distribution
\begin{equation}\label{R17}
p_k=\delta(k),\end{equation}
where $\delta(k)$ is either Kronecker symbol $\delta_{k0}$ for discrete
index $k$ or Dirac delta-function for continuous index $k$.

For standard unitary transform, one cannot find the matrix $U$ satisfying
the equation
\begin{equation}\label{R18}
U\otimes U^*={\cal L}_p.\end{equation} But if one makes averaging
with generic distribution function [not with probability
distribution (\ref{R17})], the equation
\begin{equation}\label{R19}
\langle U\otimes U^*\rangle={\cal L}_p\end{equation}
has the solution.

The standard unitary transform of density matrix is 3-parameter subset
of this 9-parameter semigroup.

Thus we constructed matrix representation of positive map of
density operators of spin-1/2 system. To construct this
representation, one needs to use the map of matrices on the
vectors discussed in the previous section. Formulas (\ref{new3})
and (\ref{new7}) can be interpreted also in the context of random
matrix representation, but we use the uniform distribution for
averaging in this case. So one has equality (\ref{new7}) in the
form
\begin{equation}\label{R20}
{\cal L}=\langle V\otimes V^*\rangle
\end{equation}
and the equality
\begin{equation}\label{R21}
\langle V^\dagger V\rangle=1,
\end{equation}
which provides constrains for used random matrices $V$.

Using random matrix formalism, the positive (but not completely
positive) maps can be presented in the form $${\cal L}=\langle
V\otimes V^*\rangle-\langle v\otimes v^*\rangle,\qquad \langle
V^\dagger V\rangle -\langle v^\dagger v\rangle=1.$$ In
\cite{Sud61} the positive maps~(\ref{new2}) and (\ref{new3}) were
used to describe non-Hamiltonian evolution of quantum states for
open systems. If the map corresponds to an extended Hamiltonian
evolution, the leading terms are of order $t^2$ (Zeno effect) and
consequently, in the $\lambda^2t$-approximation of Prigogine and
van Hove, one can derive rate equations from macroscopic equations
of motion.

We have to point out that, in general, such evolution is not
described by first-order-in-time partial differential equation.
Like in the previous case, if there are added structures of the
matrix in the form
\begin{equation}\label{eq.9}
 \rho _{id,ke}\rightarrow x_{i}y_{d}z_{k}t_{e},
 \end{equation}
which means association with the initial linear space two extra
linear spaces in which $x_{i},z_{k}$ are considered as vector
components in the $n_{1}$-dimensional linear space and $y_{d}$,
$t_{e}$ are vector components in $n_{2}$-dimensional vector space,
we will tell that one has bipartite structure of the initial space
of state [bipartite structure of the space of adjoint
representation of the group $U(n)$]. Usually the adjoint
representation of any group is defined per se without any
reference to possible substructures. Here we introduce the space
with extra structure. In addition to be space of adjoint
representation of the group $U(n)$, it has structure of bipartite
system. The generalization to multipartite ($N$-partite) structure
is straightforward. One needs only the representation of positive
integer $n^{2}$ in the form
\begin{equation}\label{eq.9a}
n^{2}=\prod_{k=1}^{N}n_{k}^{2}. \end{equation}

If one considers more general map given by superoperator (\ref{new7})
rewritten in the form
$${\cal L}=\langle V\otimes V^*\rangle,\qquad
\langle V^\dagger V\rangle=1,$$
the number of parameters determining the matrix ${\cal L}$
can be easily evaluated.
For example, if
$$
V=\left(
\begin{array}{clcr}
a&b\\
c&d\end{array}
\right),\qquad
V^*=\left(
\begin{array}{clcr}
a^*&b^*\\ c^*&d^*\end{array} \right),$$ where matrix elements are
the complex numbers, the normalization condition provides 4
constrains for the real and imaginary parts of matrix elements of
the following matrix: $$ {\cal L}=\left(
\begin{array}{clcr}
\langle |a|^2\rangle &\langle ab^*\rangle &
\langle ba^*\rangle &\langle bb^*\rangle\\
\langle ac^*\rangle &\langle aa^*\rangle &
\langle bc^*\rangle &\langle bd^*\rangle\\
\langle ca^*\rangle &\langle cb^*\rangle &
\langle da^*\rangle &\langle db^*\rangle\\
\langle cc^*\rangle &\langle cd^*\rangle &
\langle dc^*\rangle &\langle dd^*\rangle
\end{array}
\right),$$ namely, $$\langle |a|^2\rangle +\langle
|b|^2\rangle=1,\qquad \langle |c|^2\rangle +\langle |d|^2\rangle
=1,\qquad \langle a^*c\rangle +\langle a^*d\rangle=0.$$ Due to
structure of the matrix ${\cal L}$, there are 6 complex parameters
$$\langle ab^*\rangle, \quad\langle ac^*\rangle,\quad\langle
ad^*\rangle,\quad\langle bc^*\rangle,\quad\langle
bd^*\rangle,\quad\langle cd^*\rangle$$ or 12 real parameters.

Geometrical picture of positive map can be clarified if one
considers transform of the positive density matrix onto another
density matrix as transform of ellipsoid into another ellipsoid.
The generic positive transform means a generic transform of the
ellipsoid, which changes its orientation, values of semiaxis, and
position in the space. But the transform is not making from the
ellipsoid the surface like hyperboloid or paraboloid. For pure
states, the positive density matrix defines the quadratic form
which is maximally degenerated. In this sense, we say
``ellipsoid'' also including all its degenerate forms
corresponding to density matrix of ranks less than $n$ (in
$n$-dimensional case). The number of parameters defining the map
$\langle V\otimes V^*\rangle$ in $n$-dimensional case is equal to
$n^2(n^2-1)$.

\section{Local and Nonlocal Transforms}

In this section, we discuss the transforms of the density
operators of bipartite system. We concentrate on the case of two
spin-1/2 systems. Let the first system be in the state with
2$\times$2 density matrix
\begin{equation}\label{new8}
\rho_A=\left(
\begin{array}{cccc}
\rho_{11}^A& \rho_{12}^A \\
\rho_{21}^A& \rho_{22}^A
\end{array}
\right)  \end{equation} and the second system is in the state with
2$\times$2 density matrix
\begin{equation}\label{new9}
\rho_B=\left(
\begin{array}{cccc}
\rho_{11}^B& \rho_{12}^B \\
\rho_{21}^B& \rho_{22}^B
\end{array}
\right).  \end{equation} According to the suggested map, we
associate with the state density matrices (\ref{new8}) and
(\ref{new9}) the 4-vectors with components
\begin{equation}\label{new10}
\vec\rho_A=\left(
\begin{array}{c}
\rho_{11}^A\\ \rho_{12}^A \\
\rho_{21}^A\\\rho_{22}^A
\end{array}
\right),\qquad \vec\rho_B=\left(
\begin{array}{c}
\rho_{11}^B\\ \rho_{12}^B \\
\rho_{21}^B\\ \rho_{22}^B
\end{array}
\right).  \end{equation} The density matrices (\ref{new8}) and
(\ref{new9}) [i.e., vectors (\ref{new10})] belong to linear spaces
of adjoint representations of groups $U_A(2)$ and $U_B(2)$,
respectively.

For the product state (simply separable state) of composite system
with 4$\times$4 density matrix
\begin{equation}\label{new11}
\rho_{AB}=\rho_A\otimes\rho_B,
\end{equation}
the corresponding 16-vector associated to 4$\times$4 density
matrix
\begin{equation}\label{new12}
\rho_{AB}=\left(
\begin{array}{cccc}
\rho_{11}^A\rho_{11}^B& \rho_{11}^A\rho_{12}^B&
\rho_{12}^A\rho_{11}^B& \rho_{12}^A\rho_{12}^B\\
\rho_{11}^A\rho_{21}^B& \rho_{11}^A\rho_{22}^B&
\rho_{12}^A\rho_{21}^B& \rho_{12}^A\rho_{22}^B\\
\rho_{21}^A\rho_{11}^B& \rho_{21}^A\rho_{12}^B&
\rho_{22}^A\rho_{11}^B& \rho_{22}^A\rho_{12}^B\\
\rho_{21}^A\rho_{21}^B& \rho_{21}^A\rho_{22}^B&
\rho_{22}^A\rho_{21}^B& \rho_{22}^A\rho_{22}^B
\end{array}
\right)
\end{equation}
has the form
\begin{equation}\label{new13}
\vec\rho_{AB}=C\Big(\vec\rho_A\otimes\vec\rho_B\Big),
\qquad\vec\rho_A\otimes\vec\rho_B=C\vec\rho_{AB}.
\end{equation}
Here the 16$\times$16 matrix $C$ acting on the 16-component vector
$\vec\rho_A\otimes\vec\rho_B$, which is standard tensor-product of
two vectors, has the form
\begin{equation}\label{new14}
C=\left(\begin{array}{clcrclcr}
1_2&0&0&0&0&0&0&0\\0&0&1_2&0&0&0&0&0\\0&1_2&0&0&0&0&0&0\\
0&0&0&1_2&0&0&0&0\\0&0&0&0&1_2&0&0&0\\0&0&0&0&0&0&1_2&0\\
0&0&0&0&0&1_2&0&0\\0&0&0&0&0&0&0&1_2
\end{array}
\right).
\end{equation}
The matrix $C$ consists of 2$\times$2-block zero and unity
matrices.

The linear space of Hermitian matrices is equipped also by
commutator structure defining Lie algebra of the group $U(n)$. The
kernel, which defines this structure (Lie product structure) is
determined by the kernel, which determines star-product.

In the space of 16-vectors, one defines the scalar product as
follows:
\begin{equation}\label{SC1}
\vec\rho_1\cdot\vec\rho_2=\sum_{\alpha,\beta=1}^{16}\rho_{1\alpha}^*
C_{\alpha\beta}\rho_{2\beta}.
\end{equation}
The product is invariant, in view of linear transform (group
transform)
\begin{equation}\label{SC2}
\vec\rho_1\rightarrow\vec\rho_1^\prime=L\vec\rho_1\qquad
\vec\rho_2\rightarrow\vec\rho_2^\prime=L\vec\rho_2,
\end{equation}
if the 16$\times$16 matrix $L$ satisfies the condition
\begin{equation}\label{SC3}
L^{-1}=CL^\dagger C.
\end{equation}
Vector~(\ref{new13}) belongs to linear space of adjoint
representation of the unitary group $U(4)$ and star-product of the
vectors is identified with the associative algebra generated by
the Lie algebra. The local gauge transformations are defined as
tensor product of independent unitary transforms of the density
matrices $\rho_A$~(\ref{new8}) and $\rho_B$~(\ref{new9}). These
transformations are described by the group $U(2)$$\times$$U(2)$.
In the space of vectors $\vec\rho_A$ and $\vec\rho_B$, the local
transform superoperators have the matrix form
\begin{equation}\label{new15}
\hat U_A(2)\vec\rho_A=\Big(U_A(2)\otimes U_A^*(2)\Big)\vec\rho_A
\end{equation}
and
\begin{equation}\label{new16}
\hat U_B(2)\vec\rho_B=\Big(U_B(2)\otimes U_B^*(2)\Big)\vec\rho_B.
\end{equation}
In the 16-dimensional space of vectors $\vec\rho_{AB}$, the local
transforms are described by the superoperator
\begin{equation}\label{new17}
 \vec\rho _{AB}\rightarrow\vec\rho _{AB}^{\rm loc}=
 {\cal L}^{\rm loc}\vec\rho _{AB},
 \end{equation}
with the matrix
\begin{equation}\label{new18}
{\cal L}^{\rm loc}=C\Big(U_A(2)\otimes U_A^*(2)\Big)\otimes
\Big(U_B(2)\otimes U_B^*(2)\Big)C.
\end{equation}
The local positive map transforms induce in the space of adjoint
representation of the group $U(4)$ the transform of the vectors
$\vec\rho_{AB}$ associated to the matrices
\begin{equation}\label{new19}
{\cal L}_{pU}=C\Big(\sum_kp_kU^A_k\otimes U_k^{A*}\Big)\otimes
\Big(\sum_{k'}\omega_{k'}U^B_{k'}\otimes U_{k'}^{B*}\Big)C
\end{equation}
and
\begin{equation}\label{new20}
{\cal L}_{pV}=C\Big(\sum_kV^A_k\otimes V_k^{A*}\Big)\otimes
\Big(\sum_{k'}V^B_{k'}\otimes V_{k'}^{B*}\Big)C,
\end{equation}
respectively.

The matrix (\ref{new19}) can be expressed in terms of semigroup
matrices ${\cal L}_A$ and ${\cal L}_B$ as follows:
\begin{equation}\label{S2}
{\cal L}_{pU}=C\Big({\cal L}_A \otimes {\cal L}_B\Big)C,
\end{equation}
where
\begin{equation}\label{S1}
{\cal L}_A=\langle U^A \otimes U^{A^*}\rangle,\qquad{\cal L}_B=
\langle U^B\otimes U^{B^*}\rangle.
\end{equation}
Analogously
\begin{equation}\label{S3}
{\cal L}_{pV}=C\Big(\langle V^A
\otimes V^{A*}\rangle\otimes\langle V^B\otimes V^{B*}\rangle
\Big)C.
\end{equation}
The matrices ${\cal L}_{pV}$ form 18-parameter semigroup.

\section{Distributions as Vectors}

The notion of entanglement can be better clarified using the
concept of distance between the quantum states. In this section,
we consider the notion of distance between the quantum states in
terms of vectors. First, let us discuss the notion of distance
between conventional probability distributions. This notion is
well known in the classical probability theory.

Given probability distribution
$P(k)$, $k=1,2,\ldots N$, one can introduce vector $\vec{P}$ in the
form of column with components $P_{1}=P(1),$ $P_{2}=P(2),\ldots,$
$P_{N}=P(N).$  The vector satisfies the condition
\begin{equation}\label{eq.10}
\sum_{k=1}^{N}P_{k}=1. \end{equation} The set of the vectors does
not form a linear space but only a convex subset. Nevertheless, in
this set one can introduce distance between two distributions
using the vector intuition
\begin{equation}\label{eq.11}
D^{2}=\left( \vec{P}_{1}-\vec{P}_{2}\right)^{2}=\sum_{k}P_{1k}
P_{1k}+\sum_{k}P_{2k}P_{2k}-2\sum_{k}P_{1k}P_{2k}.\end{equation}

One can use another identification of distribution with vectors.

Since all $ P(k)\geq 0$, one can introduce
$\mathcal{P}_{k}=\sqrt{P(k)}$ as components of vector
$\mathcal{\vec{P}}$. The $\vec{\cal P}$ can be thought as column
with nonnegative components. Then the distance between the two
distributions takes the form
\begin{equation}\label{eq.12}
\mathcal{D}^{2}=\left(\mathcal{\vec{P}}_{1}-\mathcal{\vec{P}}_{2}\right)
^{2}=2-2\sum_{k}\sqrt{P_{1}(k)P_{2}(k)}.\end{equation}

Two different definitions (\ref{eq.10}) and (\ref{eq.11})
can be used for the notion of distance
 between the distributions.

Let us discuss now the notion of distance between the quantum
states determined by density matrices. In the density-matrix space
(in the set of linear space of adjoint $U(n)$ representation), one
can introduce distances analogously. The first case is
\begin{equation}\label{eq.13}
\mbox{Tr}\left( \rho _{1}-\rho _{2}\right)
^{2}=D^{2}\end{equation} and the second case is
\begin{equation}\label{eq.14}
\mbox{Tr}\left( \sqrt{\rho _{1}}-\sqrt{\rho _{2}}\right)
^{2}=\mathcal{D}^{2}.\end{equation} In fact, the distances
introduced can be written naturally as norms of vectors associated
to density matrices
\begin{equation}\label{new21}
D^2=|\vec \rho_1-\vec \rho_2|^2
\end{equation}
and
\begin{equation}\label{new22}
{\cal D}^2=\Big(\vec{(\sqrt \rho_1)}-\vec{(\sqrt \rho_2)}\Big)^2,
\end{equation}
respectively.

In the above expressions, we use scalar product of vectors
${\vec\rho}_1$ and ${\vec\rho}_2$ as well as scalar products of
vectors $\vec{({\sqrt \rho_1})}$ and $\vec{({\sqrt \rho_2})}$,
respectively.

Both definitions immediately follow by identification of matrices
either $\rho _{1}$ and  $\rho _{2}$ with vectors according to the
map of the previous sections or matrices $\sqrt{\rho _{1}}$ and $
\sqrt{\rho _{2}}$  with vectors. Since the density matrices $\rho
_{1}$ and $\rho _{2}$  have nonnegative eigenvalues, the matrices
$\sqrt{\rho _{1}}$ and $\sqrt{\rho _{2}}$ are defined without
ambiguity. This means that the vectors $\vec{({\sqrt \rho_1})}$
and $\vec{({\sqrt \rho_2})}$ are also defined without ambiguity.

One can easily see that the distance given as the norm of real
vector is invariant of the orthogonal group including the improper
transform, which are discrete transforms like permutations of the
vector components and changing sign of the components. The norm of
the vector is invariant with respect to all these transforms.
Thus, the invariance group of the norm contains the standard local
rotations of the orthogonal group and the discrete transforms.

\section{Invariance of Intrinsic Measure}

Given the density matrix  $\rho _{AB}$  in the linear space
equipped with bipartite structure. Then the partial traces exist
\begin{equation}\label{eq.15}
\rho _{A}=\mbox{Tr}_{B}\,\rho _{AB},
\qquad\rho_{B}=\mbox{Tr}_{A}\,\rho _{AB}.\end{equation} The
Hilbert--Schmidt distance between $\rho _{AB}$  and $\rho _{A}$
$\otimes $ $\rho _{B}$, i.e.,
\begin{equation}\label{eq.16}
e=\mbox{Tr}\left( \rho _{AB}-\rho _{A}\otimes \rho _{B}\right)
^{2} \end{equation} was considered to define the parameter $e$ as
intrinsic measure of the state entanglement. It was not proved
that this measure is invariant under local group $
U(n_{1})$$\times$$U(n_{2})$ transformations. The proof is
straightforward. One uses the structure of the tensor product in
the form
\begin{equation}\label{eq.17}
U_{n_{1}n_{2}}=U(n_{1})\otimes U(n_{2})\rightarrow \left(
\begin{array}{cccc}
U_{11}V & U_{12}V & \ldots & U_{1n}V \\
U_{21}V & U_{22}V & \ldots & U_{2n}V \\
\ldots & \ldots & \ldots & \ldots \\
U_{n1}V & U_{n2}V & \ldots & U_{nn}V
\end{array}
\right),  \end{equation} where $U_{ik}$  are matrix elements of
the $U(n_{1})$ group and $V$  is the matrix of $U(n_{2})$  group.

Due to unitarity
\begin{equation}\label{eq.18}
U_{n_{1}n_{2}}^{-1}=U_{n_{1}n_{2}}^{\dagger }\end{equation} and
using the form of this matrix given by Eq.~(\ref{eq.17}), one has
\begin{equation}\label{eq.19}
\mbox{Tr}_{B}\left( U_{n_{1}n_{2}}\rho
_{AB}U_{n_{1}n_{2}}^{\dagger }\right)
=\tilde{\rho}_{A}=U(n_{1})\rho _{A}U^{\dagger
}(n_{1}),\end{equation}
\begin{equation}\label{eq.20}\mbox{
Tr}_{A}\left( U_{n_{1}n_{2}}\rho _{AB}U_{n_{1}n_{2}}^{\dagger
}\right) =\tilde{\rho}_{B}=V\rho _{B}V^{\dagger }.\end{equation}
This means that \begin{equation}\label{eq.21}\mbox{Tr}\left( \rho
_{AB}-\rho _{A}\otimes \rho _{B}\right) ^{2}=\mbox{Tr}\left(
\tilde{\rho}_{AB}-\tilde{\rho} _{A}\otimes \tilde{\rho}_{B}
\right)^{2}. \end{equation} Thus the entanglement $e$ is invariant
under local transformations. One can introduce another measure of
entanglement, which is also invariant under local transformations
$$\widetilde
e=\left|\rho_{AB}^{1/2}-\mbox{tr}_A\,\rho_{AB}^{1/2}\,
\mbox{tr}_B\,\rho_{AB}^{1/2}\right|^2.$$ Since we have shown that
norm of a vector in adjoint representation of the unitary group is
invariant under action of the group, which is larger than the
local group, the measures of entanglement introduced are invariant
under action of local transformations.

\section{Separable Systems and Separability Criterion}

According to known definition, the system density matrix is called
separable (for composite system) if one has decomposition of the
form
\begin{equation}\label{eq.22}
\rho _{AB}=\sum_{k}p_{k}\Big(\rho _{A}^{(k)}\otimes \rho
_{B}^{(k)}\Big), \qquad \sum_{k}p_{k}=1,\qquad 1\geq p_{k}\geq
0.\end{equation} The formula does not demand orthogonality of the
density operators $\rho _{A}^{(k)}$ and $\rho _{B}^{(k)}$ for
different $k$. Since every density matrix is a convex set of pure
density matrices, one could demand that $\rho_A^{(k)}$ and
$\rho_B^{(k)}$ be pure. This formula can be interpreted in the
context of random matrix representation. In fact, one has
\begin{equation}\label{eq.22'}
\rho_{AB}=\langle\rho_A\otimes \rho_B\rangle,
\end{equation}
where $\rho_A$ and $\rho_B$ are considered as random density
matrices of the subsystems $A$ and $B$, respectively.

There are several criteria for the system to be separable. We
suggest in the next sections a new approach to the problem of
separability and entanglement based on the tomographic probability
description of quantum states. The states which cannot be
represented in the form (\ref{eq.22}) by definition are called
entangled states~\cite{J.Math.Phys:SI-Sept.2002}. Thus the states
are entangled if in formula (\ref{eq.22}) at least one coefficient
(or more)  $p_{i}$ is negative which means that the positive ones
can take values more than unity.

Let us discuss the condition for the system state to be separable.
According to Peres criterion~\cite{Peres}, the system is separable
if partial transpose of the matrix $\rho_{AB}$ (\ref{eq.22}) gives
the positive density matrix. This condition is necessary but not
sufficient. Let us discuss this condition within the framework of
positive-map matrix representation. On the example of spin-1/2
bipatite system, we have shown that the map of density matrix onto
its transpose can be included in the matrix semigroup of matrices
${\cal L}_s$. One should point out that this map cannot be
obtained by means of the averaging with all positive probability
distributions $p_k$. On the other hand, it is obvious that generic
criterion, which contains the Peres one as a partial case, can be
formulated as follows.

Let us map the density matrix $\rho_{AB}$ of a bipartite system onto vector
$\vec\rho_{AB}$. Let us act on the vector $\vec\rho_{AB}$ by an arbitrary matrix,
which represents the positive maps in subsystems A and B. Thus we get a new
vector
\begin{equation}\label{K1}
\vec\rho_{AB}^{(p)}=\Big({\cal L}_A\otimes {\cal L}_B\Big)
\vec\rho _{AB}.
\end{equation}
Let us construct the density matrix $\rho_{AB}^{(p)}$ using inverse map of
vectors onto matrices.
If the initial density matrix is separable, the new density
matrix $\rho_{AB}^{(p)}$ must be positive (and separable).

In the case of bipartite spin-1/2 system, by choosing ${\cal
L}_A=1$ and ${\cal L}_B$ being matrix coinciding with the matrix
$g^{\alpha\beta}$, we obtain the Peres criterion as a partial case
of the criterion of separability formulated above. Thus, our
criterion means that separable matrix keeps positivity under
action of tensor product of two semigroups. In the case of
bipartite spin-1/2 system, the 16$\times$16 matrix of the
semigroup tensor product is determined by 18 parameters.

Let us discuss the positive map (\ref{R20}) which is determined by
the semigroup for $n$-dimensional system. It can be realized as
follows.

The $n$$\times$$n$ Hermitian generic matrix $\rho$ is mapped onto
complex $n^2$-vector $\vec\rho$ by the map described above. The
complex vector $\vec\rho$ is mapped by means of multiplying by the
unitary matrix $S$ onto real vector $\vec\rho_{\rm r}$, i.e.,
\begin{equation}\label{G1}
\vec\rho_{\rm r}=S\vec\rho,\qquad\vec\rho=S^{-1}\vec\rho_{\rm r}.
\end{equation}
The matrix $S$ is composed from $n$ unity blocks and the blocks
\begin{equation}\label{G2}
S_b^{(jk)}=\frac {1}{\sqrt 2} \left(
\begin{array}{cccc}
1&1 \\ -i&i\end{array} \right),  \end{equation} where $j$
corresponds to column and $k$ corresponds to row in the matrix
$\rho$.

For example, in the case $n=2$, one has the vector
$\vec\rho_{\rm r}$ of the form
\begin{equation}\label{G3}
\vec\rho_{\rm r}= \left(
\begin{array}{c}
\rho_{11} \\
\sqrt 2\,\mbox{Re}\,\rho_{12}\\
\sqrt 2\,\mbox{Im}\,\rho_{12}\\
\rho_{22}
\end{array}
\right).  \end{equation}
One has the equalities
\begin{equation}\label{G4}
\vec\rho_{\rm r}^2= \vec\rho^2=\mbox{Tr}\,\rho^2.
\end{equation}
The semigroup preserves the trace of the density matrix.
Also the discrete transforms, which are described by the
matrix with diagonal matrix blocks of the form
\begin{equation}\label{G5}
{\cal D}= \left(
\begin{array}{clcr}
1&0&0&0 \\
0&1&0&0 \\
0&0&-1&0 \\
0&0&0&1
\end{array}
\right),  \end{equation} preserve positivity of the density
matrix.

For the spin case, the semigroup contains 12 parameters.

Thus, the direct product of the semigroup (\ref{R20}) and the
discrete group of the transform $D$ is the positive map preserving
positivity of the density operator.

\section{Symbols, Star-Product and Entanglement}

In this section, we describe how entangled states and separable
states can be considered using properties of symbols and density
operators of different kinds, e.g., from the viewpoint of Wigner
function or tomogram. The general scheme  of constructing the
operator symbols is as follows~\cite{Marmo}.

Given a Hilbert space $H$ and an operator $\hat A $ acting on this
space, let us suppose that we have a set of operators $\hat U({\bf
x})$ acting on $H$, a $n$-dimensional vector ${\bf
x}=(x_1,x_2,\ldots,x_n)$ labels the particular operator in the
set. We construct the $c$-number function $f_{\hat A}({\bf x})$
(we call it the symbol of operator $\hat A$) using the definition
\begin{equation}\label{eq.1b}
f_{\hat A}({\bf x})=\mbox{Tr}\left[\hat A\hat U({\bf x})\right].
\end{equation}
Let us suppose that relation~(\ref{eq.1b}) has an inverse, i.e.,
there exists a set of operators $\hat D({\bf x})$ acting on the
Hilbert space such that
\begin{equation}\label{eq.2b}
\hat A= \int f_{\hat A}({\bf x})\hat D({\bf x})~d{\bf x}, \qquad
\mbox{Tr}\, \hat A= \int f_{\hat A}({\bf x})\,\mbox{Tr}\, \hat
D({\bf x})~d{\bf x}.
\end{equation}
Then, we will consider relations~(\ref{eq.1b}) and~(\ref{eq.2b})
as relations determining the invertible map from the operator
$\hat A$ onto function $f_{\hat A}({\bf x})$. Multiplying both
sides of Eq.~(\ref{eq.2}) by the operator $\hat U({\bf x}')$ and
taking trace, one has the consistency condition satisfied for the
operators $\hat U({\bf x}')$ and $\hat D({\bf x})$
\begin{equation}\label{eq.2b'}
\mbox{Tr}\left[\hat U({\bf x}')\hat D({\bf x})\right]
=\delta\left({\bf x}'-{\bf x}\right).
\end{equation}
The consistency condition~(\ref{eq.2b'}) follows from the relation
\begin{equation}\label{eq.2aa}
f_{\hat A}({\bf x})=\int K({\bf x}, {\bf x}')f_{\hat A}({\bf x}')
\,d{\bf x}'.
\end{equation}
The kernel in~(\ref{eq.2aa}) is equal to the standard Dirac
delta-function, if the set of functions $f_{\hat A}({\bf x})$ is a
complete set.

In fact, we could consider relations of the form
\begin{equation}\label{eq.3b}
\hat A\rightarrow f_{\hat A}({\bf x})
\end{equation}
and
\begin{equation}\label{eq.4b}
f_{\hat A}({\bf x})\rightarrow\hat A.
\end{equation}
The most important property of the map is the existence of
associative product (star-product) of functions.

We introduce the product (star-product) of two functions $f_{\hat
A}({\bf x})$ and $f_{\hat B}({\bf x})$ corresponding to two
operators $\hat A$ and $\hat B$ by the relationships
\begin{equation}\label{eq.5b}
f_{\hat A\hat B}({\bf x})=f_{\hat A}({\bf x})*
f_{\hat B} ({\bf x}):=\mbox{Tr}\left[\hat A\hat B\hat U({\bf x})
\right].
\end{equation}
Since the standard product of operators on a Hilbert space is an
associative product, i.e., $\hat A(\hat B \hat C)=(\hat A\hat
B)\hat C$, it is obvious that formula~(\ref{eq.5b}) defines an
associative product for the functions $f_{\hat A}({\bf x})$, i.e.,
\begin{equation}\label{eq.6b}
f_{\hat A}({\bf x})*\Big(f_{\hat B}({\bf x})
*f_{\hat C}({\bf x})\Big)=
\Big(f_{\hat A}({\bf x})*f_{\hat B}({\bf x})\Big)
*f_{\hat C}({\bf x}).
\end{equation}

Using formulas~(\ref{eq.1b}) and (\ref{eq.2b}), one can write down
a composition rule for two symbols $f_{\hat A}({\bf x})$ and
$f_{\hat B}({\bf x})$, which determines star-product of these
symbols. The composition rule is described by the formula
\begin{equation}\label{eq.25b}
f_{\hat A}({\bf x})*f_{\hat B}({\bf x})=
\int f_{\hat A}({\bf x}'')f_{\hat B}({\bf x}')
K({\bf x}'',{\bf x}',{\bf x})\,d{\bf x}'\,d{\bf x}''.
\end{equation}
The kernel in the integral of (\ref{eq.25b}) is determined by the
trace of product of the basic operators, which we use to construct
the map
\begin{equation}\label{eq.26b}
K({\bf x}'',{\bf x}',{\bf x})=
\mbox{Tr}\left[\hat D({\bf x}'')\hat D({\bf x}')
\hat U({\bf x})\right].
\end{equation}
Formula~(\ref{eq.26b}) can be extended to the case of star-product
of $N$ symbols of operators $\hat A_1,\hat A_2,\ldots,\hat A_N$.
Thus one has
\begin{eqnarray}\label{eq.26'}
W_{\hat A_1}({\bf x})*W_{\hat A_2}({\bf x})*\cdots * W_{\hat
A_N}({\bf x})&=&\int W_{\hat A_1}({\bf x}_1) W_{\hat A_2}({\bf
x}_2)\cdots W_{\hat A_N}({\bf x}_N)\nonumber\\ &&\times
K\left({\bf x}_1,{\bf x}_2,\ldots,{\bf x}_N,{\bf x}\right) \,d{\bf
x}_1\,d{\bf x}_2\cdots \,d{\bf x}_N,
\end{eqnarray}
where the kernel has the form
\begin{equation}\label{eq.26''}
K\left({\bf x}_1,{\bf x}_2,\ldots,{\bf x}_N,{\bf x}\right)=
\mbox{Tr}\left[\hat D({\bf x}_1)\hat D({\bf x}_2)
\cdots\hat D({\bf x}_N)\hat U({\bf x})\right].
\end{equation}
Since this kernel determines the associative star-product of $N$
symbols, it can be expressed in terms of the kernel of
star-product of two symbols. The trace of an operator $\hat A^{N}$
is determined by the kernel as follows:
\begin{eqnarray}\label{eq.26'''}
\mbox{Tr}\, \hat A^N&=&\int W_{\hat A}({\bf x}_1) W_{\hat A}({\bf
x}_2)\cdots W_{\hat A}({\bf x}_N) \nonumber\\ &&\times
\mbox{Tr}\left[\hat D({\bf x}_1)\hat D({\bf x}_2) \cdots\hat
D({\bf x}_N)\right] \,d{\bf x}_1\,d{\bf x}_2\cdots \,d{\bf x}_N.
\end{eqnarray}
When the operator $\hat A$ is a density operator of a quantum
state, formula~(\ref{eq.26'''}) determines the generalized purity
parameter of the state. When the operator $\hat A$ is equal to the
product of two density operators and $N=1$, formula
(\ref{eq.26'''}) determines the fidelity.

\section{Weyl Symbol}

In this section, we will consider a known
example of the Heisenberg--Weyl-group representation.
As operator $\hat U({\bf x})$, we take the Fourier
transform of displacement operator
$\hat D(\xi)$
\begin{equation}\label{eq.27b}
\hat U({\bf x})=\int \exp\left(\frac{x_1+ix_2}{\sqrt 2}
\mbox{\boldmath$\xi$}^*-\frac{x_1-i x_2}{\sqrt 2}
\mbox{\boldmath$\xi$}\right) \hat
D(\mbox{\boldmath$\xi$})\pi^{-1}~d^2 \mbox{\boldmath$\xi$},
\end{equation}
where $\mbox{\boldmath$\xi$}$ is a complex number,
$\mbox{\boldmath$\xi$}=\xi_1+i\xi_2$, and the vector ${\bf
x}=(x_1,x_2)$ can be considered as ${\bf x}=(q,p)$, with $q$ and
$p$ being the position and momentum. One can see that
$$\mbox{Tr}\,\hat U({\bf x})=1.$$ The displacement operator may be
expressed through the creation and annihilation operators in the
form
\begin{equation}\label{eq.28b}
\hat D(\mbox{\boldmath$\xi$})=\exp(\mbox{\boldmath$\xi$}
\hat a^\dagger-\mbox{\boldmath$\xi$}^*\hat a).
\end{equation}
The displacement operator is used to create coherent states from
the vacuum state. For the creation and annihilation operators, one
has
\begin{equation}\label{eq.29b}
\hat a=\frac{\hat q+i\hat p}{\sqrt 2}, \qquad \hat
a^\dagger=\frac{\hat q- i\hat p}{\sqrt 2},
\end{equation}
where $\hat q$ and $\hat p$ may be thought as the coordinate and
momentum  operators for the carrier space of an harmonic
oscillator. The operator $\hat a$ and its Hermitian conjugate
$\hat a^\dagger$ satisfy the boson commutation relation $$[\hat
a,\hat a^\dagger]=\hat {\bf 1}.$$

Let us introduce the Weyl symbol for an arbitrary operator $\hat
A$ using the definition given by Eq.~(\ref{eq.1b})
\begin{equation}\label{eq.30b}
W_{\hat A}({\bf x})=\mbox{Tr}\left[\hat A\hat U({\bf x}) \right].
\end{equation}
The form of operator $\hat U({\bf x})$ is given by
Eq.~(\ref{eq.27b}). One can check that Weyl symbols of the
identity operator $\hat {\bf 1}$, position operator $\hat q$ and
momentum operator $\hat p$ have the form
\begin{equation}\label{eq.30'}
W_{\hat {\bf 1}}(q,p)=1,\qquad W_{\hat q}(q,p)=q,\qquad W_{\hat
p}(q,p)=p.
\end{equation}
The inverse transform, which expresses the operator $\hat A$
through its Weyl symbol, is of the form
\begin{equation}\label{eq.31b}
\hat A=\int
W_{\hat A}({\bf x})\hat U({\bf x})\,\frac{d{\bf x}}{2\pi}\,.
\end{equation}
One can check that for $W_{\hat {\bf 1}}({\bf x})=1$,
formula~(\ref{eq.31b}) reproduces the identity operator, i.e.,
\begin{equation}\label{eq.31'}
\int \hat U({\bf x})\,\frac{d{\bf x}}{2\pi}=\hat {\bf 1}.
\end{equation}
Comparing (\ref{eq.31b}) with~(\ref{eq.2b}), one can see that the
operator $\hat D({\bf x})$ is connected with $\hat U({\bf x})$ by
the relationship
\begin{equation}\label{eq.32b}
\hat D({\bf x})=\frac{\hat U({\bf x})}{2\pi}\,.
\end{equation}
Let us consider now star-product of two Weyl symbols (it is
usually called Moyal star-product). If one takes two operators
$\hat A_1$ and $\hat A_2$, which are expressed through Weyl
symbols by formulas
\begin{equation}\label{eq.33b}
 \hat A_1=\int
W_{\hat A_1}({\bf x}')\hat U({\bf x}')\,\frac{d{\bf x}'}{2\pi}\,,
\qquad \hat A_2=\int W_{\hat A_2}({\bf x}'') \hat U({\bf
x}'')\,\frac{d{\bf x}''}{2\pi}\,,
\end{equation}
with vectors $${\bf x}'=(x_1',x_2')\qquad\mbox{and}\qquad {\bf
x}''=(x_1'',x_2''),$$ the operator $\hat A$ (product of operators
$\hat A_1$ and $\hat A_2)$ has Weyl symbol given by
\begin{eqnarray}
 W_{\hat A}({\bf x})=\mbox{Tr}\,
\Big[\hat A\hat U({\bf x})\Big]=\frac{1}{4\pi^5}\,\int
d{\bf x}'\,d{\bf x}''\,d^2{\bf\xi}\,d^2{\bf\xi}'\,d^2{\bf\xi}''\,
W_{\hat A_1}({\bf x}')W_{\hat A_2}({\bf x}'')\nonumber\\
 \times\exp\Big\{2^{-1/2}\Big[(\xi_1'-i\xi_2')(x_1'+ix_2')-
(\xi_1'+i\xi_2')(x_1'-ix_2')\nonumber\\
 +(\xi_1''-i\xi_2'')(x_1''+ix_2'')
-(\xi_1''+i\xi_2'')(x_1''-ix_2'')+
(\xi_1-i\xi_2)(x_1+ix_2)\nonumber\\
 -(\xi_1+i\xi_2)(x_1-ix_2)\Big]\Big\}\,
\mbox{Tr}\left[\hat D(\mbox{\boldmath$\xi$}') \hat
D(\mbox{\boldmath$\xi$}'') \hat D(\mbox{\boldmath$\xi$})\right],
\label{eq.34b}
\end{eqnarray}
where $\mbox{\boldmath$\xi$}
=\xi_1+i\xi_2$, with $~\mbox{\boldmath$\xi$}'=\xi_1'+i\xi_2'$
and $~\mbox{\boldmath$\xi$}''=\xi_1''+i\xi_2''$.

Using properties of displacement operators
\begin{equation}\label{eq.35b}
 \hat D(\mbox{\boldmath$\xi$}')\hat D
(\mbox{\boldmath$\xi$}'')= \hat
D(\mbox{\boldmath$\xi$}'+\mbox{\boldmath$\xi$}'') \exp\Big(i\mbox{
Im}\,(\mbox{\boldmath$\xi$}' \mbox{\boldmath$\xi$}''^*)\Big),
\qquad \mbox{Tr}\left[\hat D(\mbox{\boldmath$\xi$})\right]
=\pi\delta^2(\mbox{\boldmath$\xi$}),
\end{equation}
one can get known explicit form of the kernel, which determines
star-product of Weyl symbols.Thus we described the construction of
Weyl symbols, including Wigner function, by means of the
star-product formalism.

\section{Tomographic Representation}

In this section, we will consider an example of the probability
representation of quantum mechanics~\cite{JOB}. In the probability
representation of quantum mechanics, the state is described by a
family of probabilities~[22--24].
According to the general scheme, one can introduce for the
operator $\hat A$ the function $f_{\hat A}({\bf x})$, where $${\bf
x}=(x_1,x_2,x_3)\equiv (X,\mu,\nu),$$ which we denote here as
$w_{\hat A}(X,\mu,\nu)$ depending on the position $X$ and the
parameters $\mu$ and $\nu$ of the reference frame
\begin{equation}\label{eq.53}
w_{\hat A}(X,\mu,\nu)=\mbox{Tr}\left[\hat A
\hat U({\bf x})\right].
\end{equation}
We call the function $w_{\hat A}(X,\mu,\nu)$ the tomographic
symbol of the operator $\hat A$. The operator $\hat U(x)$ is
given by
\begin{eqnarray}\label{eq.54}
 \hat U({\bf x})\equiv \hat U(X,\mu,\nu)&=&
\exp\left(\frac{i\lambda}{2}\left(\hat q\hat p +\hat p\hat
q\right)\right) \exp\left(\frac{i\theta}{2}\left(\hat q^2 +\hat
p^2\right)\right) \mid X\rangle\langle X\mid\nonumber\\ &&
\times\exp\left(-\frac{i\theta}{2}\left(\hat q^2 +\hat
p^2\right)\right) \exp\left(-\frac{i\lambda}{2}\left(\hat q\hat p
+\hat p\hat q\right)\right)\nonumber\\&=&\hat U_{\mu\nu}\mid
X\rangle\langle X\mid\hat U_{\mu\nu}^\dagger.
\end{eqnarray}
The angle $\theta$ and parameter $\lambda$ in terms of the
reference frame parameters are given by $$
\mu=e^{\lambda}\cos\theta, \qquad \nu=e^{-\lambda}\sin\theta. $$
Moreover, $\hat q$ and $\hat p$ are position and momentum
operators
\begin{equation}\label{eq.54'}
\hat q\mid X\rangle=X\mid X\rangle
\end{equation}
and $\mid X\rangle\langle X\mid$ is the projection density. One
has the canonical transform of quadratures $$\hat X=\hat
U_{\mu\nu}\,\hat q\,\hat U^{\dagger}_{\mu\nu} =\mu \hat q+\nu\hat
p,$$ $$ \hat P=\hat U_{\mu\nu}\,\hat p\,\hat U^{\dagger}_{\mu\nu}=
\frac{1+\sqrt{1-4\mu^2\nu^2}}{2\mu}\,\hat p-
\frac{1-\sqrt{1-4\mu^2\nu^2}}{2\nu}\,\hat q. $$

Using the approach of \cite{MendesJPA} one can obtain the
relationship $$\hat U(X,\mu,\nu)=\delta(X-\mu \hat q-\nu\hat p).$$
In the case we are considering, the inverse transform determining
the operator in terms of tomogram [see Eq.~(\ref{eq.2b})] will be
of the form
\begin{equation}\label{eq.55}
\hat A=\int w_{\hat A}(X,\mu,\nu) \hat
D(X,\mu,\nu)\,dX\,d\mu\,d\nu,
\end{equation}
where
\begin{equation}\label{eq.56}
\hat D({\bf x})\equiv\hat D(X,\mu,\nu)=\frac{1}{2\pi}
\exp\left(iX-i\nu\hat p-i\mu\hat q\right),
\end{equation}
i.e.,
\begin{equation}\label{eq.56q}
\hat D(X,\mu,\nu)
=\frac{1}{2\pi}\exp(iX)\hat D\Big(\mbox{\boldmath$\xi$}
(\mu,\nu)\Big).
\end{equation}
The unitary displacement operator in (\ref{eq.56q}) reads now $$
\hat D\Big( \mbox{\boldmath$\xi$}(\mu,\nu)\Big)
=\exp\Big(\mbox{\boldmath$\xi$}(\mu,\nu) \hat
a^+-{\mbox{\boldmath$\xi$}}^*(\mu,\nu)\hat a\Big), $$ where
$\mbox{\boldmath$\xi$}(\mu,\nu)=\xi_1+i\xi_2$, with
$\xi_1=\mbox{Re}\, (\mbox{\boldmath$\xi$})={\nu}/{\sqrt2}$
 and $\xi_2=\mbox{Im}\,(\mbox{\boldmath$\xi$})
=-{\mu}/{\sqrt2}$.

Trace of the above operator which provides the kernel
determining the trace of an arbitrary operator in the
tomographic representation reads
$$\mbox{Tr}\,\hat D({\bf x})=
e^{iX}\delta (\mu)\delta(\nu).$$
The function $w_{\hat A}(X,\mu,\nu)$ satisfies the relation
\begin{equation}\label{eq.56'}
w_{\hat A}\left(\lambda X,\lambda \mu,\lambda\nu\right)
=\frac{1}{|\lambda|}\,w_{\hat A}(X,\mu,\nu).
\end{equation}
This means that the tomographic symbols of operators are homogeneous
functions of three variables.

If one takes two operators $\hat A_1$ and $\hat A_2$, which are
expressed through the corresponding functions by the formulas
\begin{eqnarray}
\hat A_1&=&\int w_{\hat A_1}(X',\mu',\nu')\hat
D(X',\mu',\nu')\,dX'\,d\mu' \,d\nu', \nonumber\\ &&\label{eq.57}\\
\hat A_2&=&\int w_{\hat A_2}(X'',\mu'',\nu'') \hat
D(X'',\mu'',\nu'')dX''\,d\mu''\,d\nu'', \nonumber
\end{eqnarray}
and $\hat A$ denotes the product of $\hat A_1$ and $\hat A_2$,
then the function $w_{\hat A}(X,\mu,\nu)$, which corresponds to
$\hat A$, is star-product of functions $w_{\hat A_1}(X,\mu,\nu)$
and $w_{\hat A_2}(X,\mu,\nu)$. Thus this product $$ w_{\hat
A}(X,\mu,\nu)=w_{\hat A_1}(X,\mu,\nu) *w_{\hat A_2}(X,\mu,\nu) $$
reads
\begin{equation}\label{eq.58}
w_{\hat A}(X,\mu,\nu)=\int w_{\hat A_1}({\bf x}'') w_{\hat
A_2}({\bf x}')K({\bf x}'',{\bf x}', {\bf x})\,d{\bf x''}\,d{\bf
x'},
\end{equation}
with kernel given by
\begin{equation}\label{eq.59}
K({\bf x}'',{\bf x}',{\bf x})=
\mbox{Tr}\left[\hat D(X'',\mu'',\nu'')
\hat D(X',\mu',\nu')\hat U(X,\mu,\nu)\right].
\end{equation}
The explicit form of the kernel reads
\begin{eqnarray}\label{KERNEL}
 &&K(X_1,\mu_1,\nu_1,X_2,\mu_2,\nu_2,X\mu,\nu)\nonumber\\
 &&=\frac{\delta\Big(\mu(\nu_1+\nu_2)-\nu(\mu_1+\mu_2)\Big)}{4\pi^2}
\,\exp\left(\frac{i}{2}\Big\{\left(\nu_1\mu_2-\nu_2\mu_1\right)
+2X_1+2X_2\right.\nonumber\\ && \left.\left.
-\left[\frac{1-\sqrt{1-4\mu^2\nu^2}}{\nu}
\left(\nu_1+\nu_2\right)+\frac{1+\sqrt{1-4\nu^2\mu^2}}{\mu}
\left(\mu_1+\mu_2\right) \right]X\right\}\right).
\end{eqnarray}
The kernel for star-product of $N$ operators is
\begin{eqnarray}\label{KERNELSTAR}
&&K\left(X_1,\mu_1,\nu_1,X_2,\mu_2,\nu_2,\ldots,
X_N,\mu_N,\nu_N,X,\mu,\nu\right)\nonumber\\ &&
=\frac{\delta\left(\mu\sum_{j=1}^N\nu_j-\nu\sum_{j=1}^N\mu_j\right)}
{(2\pi)^N}\,\exp\left(\frac{i}{2}\,\left\{\sum_{k<j=1}^N
\left(\nu_k\mu_j-\nu_j\mu_k\right)+2\sum_{j=1}^NX_j\right.\right.
\nonumber\\&&
 \left.\left.-\left[
\frac{1-\sqrt{1-4\mu^2\nu^2}}{\nu}
\left(\sum_{j=1}^N\nu_j\right)+
 \frac{1+\sqrt{1-4\mu^2\nu^2}}{\mu}\left(\sum_{j=1}^N\mu_j\right)
\right]X\right\}\right).
\end{eqnarray}
The above kernel can be expressed in terms of the kernel
determining star-product of two operators.

\section{Multipartite Systems}

Let us assume that for multimode ($N$-mode) system one has
\begin{eqnarray}
\hat U(\vec y)=\prod_{k=1}^N\hat U\left(\vec
x^{(k)}\right),\label{P6}\\ \hat D(\vec y)=\prod_{k=1}^N\hat
D\left(\vec x^{(k)}\right),\label{P7}
\end{eqnarray}
where
\begin{equation}\label{P8}
\vec y=\Big(x_1^{(1)},x_2^{(1)},\ldots,x_m^{(1)},x_1^{(2)},x_2^{(2)},
\ldots,x_m^{(N)}\Big).
\end{equation}
This means that symbol of density operator of the composite system
reads
\begin{equation}\label{P9}
f_\rho(\vec y)=\mbox{Tr}\,\Big[\hat\rho\prod_{k=1}^N
\hat U(\vec x^{(k)})\Big].\end{equation}
The inverse transform reads
\begin{equation}\label{P10}
\hat\rho=\int d\vec y\,f_\rho(\vec y)\prod_{k=1}^N \hat D(\vec
x^{(k)}),\qquad d\vec y=\prod_{k=1}^N\prod_{s=1}^mdx_s^{(k)}.
\end{equation}
If the symbol corresponds to a system Wigner function $W(\vec
q,\vec p)$, the operator $\hat U(\vec x^{(k)})$, where $\vec
x^{(k)}=(q_k,p_k)$, was discussed above. It has the form
\begin{equation}\label{P11}
\hat U(\vec x^{(k)})={\cal D}(\alpha _k)(-1)^{a_k^\dagger a_k}
{\cal D}(-\alpha _k),
\end{equation}
where
\begin{equation}\label{P12}
\alpha_k=\frac{1}{\sqrt 2}\,(q_k+ip_k),\qquad {\cal D}(\alpha_k)=
e^{\alpha_ka_k^\dagger-\alpha_k^*a_k},\qquad [a_k,a_m]=0,\qquad
[a_k,a_m^\dagger]=\delta_{km}.\end{equation} The operator
\begin{equation}\label{P13}
\hat D(\vec x^k)=\frac {1}{\pi}\,\hat U(\vec
x^{(k)}).\end{equation} The tomographic symbols are also defined
by analogous formulas with specific $\vec x$ and operators $\hat
U(\vec x_k)$, $\hat D(\vec x_k)$.

Now we formulate properties of symbols in the case of entangled and
separable states, respectively.

Given a composite $m$-partite system with density operator
$\hat\rho$.

If the nonnegative operator can be presented in the form of
`probabilistic sum'
\begin{equation}\label{P14}
\hat\rho=\sum_{\vec z}{\cal P}(\vec z)\hat\rho_{\vec
z}^{(a_1)}\otimes \hat\rho_{\vec z}^{(a_2)}\otimes\cdots\otimes
\hat\rho_{\vec z}^{(a_m)},\end{equation} with positive probability
distribution function ${\cal P}(\vec z)$, where components of
$\vec z$ can be either discrete or continuous, we call the state
`separable state'. This means that the symbol of the state can be
presented in the form
\begin{equation}\label{P15}
f_\rho(\vec y)=\sum_{\vec z}{\cal P}(\vec
z)\prod_{k=1}^mf_\rho^{(a_k)} (\vec x_{k},\vec z).\end{equation}
For example, the Wigner function of separable state of bipartite
system has the form
\begin{equation}\label{P16}
W(q_1,q_2,p_1,p_2)=\sum_{\vec z}{\cal P}(\vec z)W^{(1)}(q_1,p_1,\vec z)
W^{(2)}(q_2,p_2,\vec z).
\end{equation}
Analogous formula can be written for the tomogram of separable
state.

\section{Spin Tomography}

 Below we concentrate on bipartite spin systems.

The tomographic probability (spin tomogram) completely determines
the density matrix of a spin state. It has been introduced in
\cite{Dod,Olga,Marmo}.

The tomographic probability for spin-$j$  state is defined via the
density matrix by the formula
\begin{equation}\label{eq.23c}
\langle jm\mid D^{\dagger }(g)\rho D(g)\mid
jm\rangle=W^{(j)}(m,\vec{0}), \qquad m=-j,-j+1,\ldots,j,
\end{equation} where $D(g)$ is the matrix of $SU(2)$-group
representation depending on the group element $g$ determined by
three Euler angles. The set of the tomogram values for each $\vec
0$ is an overcomplete set. We need only finite number of
independent locations which will give information on the density
matrix of the spin state.
 Due to structure of the formula, there are only two
Euler angles involved. They are combined into the unit
vector
\begin{equation}\label{eq.24c} \vec{0}=(\cos \phi \sin
\vartheta ,\sin \phi \sin \vartheta ,\cos \vartheta
).\end{equation} This is the map from $S^3$ to $S^2$.

The physical meaning of the probability $W(m,\vec{0})$  is the
following.

It is the probability to find, in the state with the density
matrix $\rho $, the spin projection on direction  $\vec{0}$ equal
to $m$. For bipartite system, the tomogram is defined as follows:
\begin{equation}\label{eq.25c}
W(m_{1}m_{2}\vec{0}_{1}\vec{0}_{2})=\langle
j_{1}m_{1}j_{2}m_{2}\mid D^{\dagger }(g_{1})D^{\dagger
}(g_{2})\rho D(g_{1})D(g_{2})\mid j_{1}m_{1}j_{2}m_{2}\rangle.
\end{equation} It completely determines the density matrix $\rho
$. It has the meaning of joint probability distribution for spin
$j_{1}$ and $j_{2}$ projections $m_{1}$ and $m_{2}$  on directions
$ \vec{0}_{1}$ and $\vec{0}_{2}.$ Since the map $\rho
\rightleftharpoons W$ is linear and invertible, the definition of
separable system can be rewritten in the following form of
decomposition of the joint probability into sum of products (of
factorized probabilities):
\begin{equation}\label{eq.26c}
W(m_{1}m_{2}\vec{0}_{1}\vec{0}_{2})=
\sum_{k}p_{k}W^{(k)}(m_{1}\vec{0}_{1})
\tilde{W}^{(k)}(m_{2}\vec{0}_{2}).\end{equation} This form can be
considered to formulate the criterion of separability of the two
spin state.

The state is separable iff the tomogram can be written in the form
(\ref{eq.26c}) with $\sum_{k}p_{k}=1$,  $p_{k}\geq 0.$ It seems
that we simply use the definition but, in fact, we cast the
problem of separability into the form of property of the positive
joint probability distribution of two random variables. This is
area of probability theory and one can use results and theorems on
the joint probability distributions. If one does not use any
theorem, one has to study solvability of relation (\ref{eq.26c})
considered as the equation for unknown probability distribution
$p_{k}$ and unknown probability functions
$W^{(k)}(m_{1}\vec{0}_{1})$ and $W^{(k)}(m_{2}\vec{0}_{2}).$

\section{Example of Spin-${1}/{2}$  Bipartite System}

For spin-${1}/{2}$ state, the generic density matrix can be
presented in the form
\begin{equation}\label{eq.27c}
\rho =\frac{1}{2}\left( 1+\vec{\sigma}\cdot \vec{n}\right), \qquad
\vec{n}=(n_{1},n_{2},n_{3}), \end{equation} where $\vec{\sigma}$
are Pauli matrices and $\vec{n}^{2}\leq 1,$  with vector $\vec{n}$
for a pure state being unit vector. This decomposition means that
we use as basis in 4-dimensional vector space the vectors
corresponding to Pauli matrices and unit matrix, i.e.,
\begin{equation}\label{P1}
\vec\sigma_1= \left(
\begin{array}{c}
0 \\
1\\
1\\
0\end{array}
\right),\qquad
\vec\sigma_2= \left(
\begin{array}{c}
0 \\
-i\\
i\\
0\end{array}
\right),\qquad
\vec\sigma_3= \left(
\begin{array}{c}
1\\
0\\
0\\
-1\end{array}
\right),\qquad
\vec 1= \left(
\begin{array}{c}
1\\
0\\
0\\
1\end{array}
\right).  \end{equation}
The density matrix vector
\begin{equation}\label{P2}
\vec\rho= \left(
\begin{array}{c}
\rho_{11} \\
\rho_{12}\\
\rho_{21}\\
\rho_{22}\end{array}
\right)\end{equation}
is decomposed in terms of the basis vectors
\begin{equation}\label{P3}
\vec\rho=\frac {1}{2}\,\Big(\vec 1+n_1\vec\sigma_1+n_2\vec\sigma_2
+n_3\vec\sigma_3\Big).
\end{equation}
It means that tomogram of spin-1/2 state can be given in the form
\begin{equation}\label{eq.28c}
W\left(\frac{1}{2},\vec{0}\right)=\left(
\frac{1}{2}+\frac{\vec{n}\cdot \vec{0}}{2}\right), \qquad
W\left(-\frac{1}{2},\vec{0}\right)=\left(
\frac{1}{2}-\frac{\vec{n}\cdot \vec{0}}{2}\right). \end{equation}
Inserting these probability values into relation (\ref{eq.26c})
for each value of $k$ we get the relationships:
\begin{equation}\label{eq.29c}
W\left(\frac{1}{2},\frac{1}{2},\vec{0}_{1},\vec{0}_{2}\right)
=\frac{1}{4}+\frac{1}{2}\left( \sum_{k}p_{k}\vec{n}_{k}\right)
\cdot \vec{0}_{1}+\frac{1}{2}\left(
\sum_{k}p_{k}\vec{n}_{k}^*\right) \cdot
\vec{0}_{2}+\sum_{k}p_{k}\left( \vec{n}_{k}\cdot
\vec{0}_{1}\right) \left( \vec{n}_{k}^8\cdot \vec{0}_{2}\right),
\end{equation}
\begin{equation}\label{eq.30c}
W\left(\frac{1}{2},-\frac{1}{2},\vec{0}_{1},\vec{0}_{2}\right)
=\frac{1}{4}+\frac{1}{2}\left( \sum_{k}p_{k}\vec{n}_{k}\right)
\cdot \vec{0}_{1}-\frac{1}{2}\left(
\sum_{k}p_{k}\vec{n}_{k}^*\right) \cdot \vec{0}_{2}-
\sum_{k}p_{k}\left( \vec{n}_{k}\cdot \vec{0}_{1}\right) \left(
\vec{n}_{k}^*\cdot \vec{0}_{2}\right), \end{equation}
\begin{equation}\label{eq.31c}
W\left(-\frac{1}{2},\frac{1}{2},\vec{0}_{1},\vec{0}_{2}\right)
=\frac{1}{4}-\frac{1}{2}\left( \sum_{k}p_{k}\vec{n}_{k}\right)
\cdot \vec{0}_{1}+\frac{1}{2}\left(
\sum_{k}p_{k}\vec{n}_{k}^*\right) \cdot \vec{0}_{2}-\sum_{k}p_{k}
\left( \vec{n}_{k}\cdot \vec{0}_{1}\right) \left(
\vec{n}_{k}^*\cdot \vec{0}_{2}\right). \end{equation}
 One has the normalization property
\begin{equation}\label{eq.32c}
\sum_{m_{1},\,m_{2}=-{1}/{2}}^{{1}/{2}}W(m_{1}m_{2}
\vec{0}_{1}\vec{0}_{2})=1.\end{equation} One easily gets
\begin{equation}\label{eq.33c}
W\left(\frac{1}{2},\frac{1}{2},\vec{0}_{1},\vec{0}_{2}\right)
+W\left(\frac{1}{2},-\frac{1}{2},\vec{0}_{1},\vec{0}_{2}\right)=
\frac{1}{2}+\left( \sum_{k}p_{k}\vec{n}_{k}\right) \cdot
\vec{0}_{1}.\end{equation} This means that derivative in
$\vec{0}_{1}$ on the left-hand side gives
\begin{equation}\label{eq.34c}
\frac{\partial }{\partial \vec{0}_{1}}\left[
W\left(\frac{1}{2},\frac{1}{2},
\vec{0}_{1},\vec{0}_{2}\right)+W\left(\frac{1}{2},-\frac{1}{2},\vec{0}_{1},
\vec{0}_{2}\right)\right] =\left(
\sum_{k}p_{k}\vec{n}_{k}\right).\end{equation} Analogously
\begin{equation}\label{eq.35c}
\frac{\partial }{\partial \vec{0}_{2}}\left[
W\left(\frac{1}{2},\frac{1}{2},
\vec{0}_{1},\vec{0}_{2}\right)+W\left(-\frac{1}{2},\frac{1}{2},\vec{0}_{1},
\vec{0}_{2}\right)\right] =\left( \sum_{k}p_{k}\vec{n}_{k}^{\left(
\star \right) }\right). \end{equation} Taking the sum of
(\ref{eq.30c}) and (\ref{eq.31c})) one sees that
\begin{equation}\label{eq.36c}
\frac{1}{2} \frac{\partial }{\partial \vec{0}_{i}}\frac{\partial
}{\partial \vec{0}_{j}}\left[
W\left(\frac{1}{2},-\frac{1}{2},\vec{0}_{1},\vec{0}_{2}\right)
+W\left(-\frac{1}{2},\frac{1}{2},\vec{0}_{1},\vec{0}_{2}\right)\right]
=- \sum_{k}p_{k}(n_{k})_{i}(n_{k}^{\left( \star \right) })_{j}.
\end{equation}

Since we look for solution where $p_{k}\geq 0$, we can introduce
\begin{equation}\label{eq.37c}
\vec{N}_{k}=\sqrt{p_{k}}\vec{n}_{k}, \quad \vec{N}_{k}^{\left(
\star \right) }=\sqrt{p_{k}}\vec{n}_{k}^{\left( \star \right) }.
\end{equation} This means that the derivative in (\ref{eq.36c}) can be
presented as tensor \begin{equation}\label{eq.38c}
-T_{ij}=\sum_{k}(N_{k})_{i}(N_{k}^{\left( \star \right) })_j.
\end{equation}
One has \begin{equation}\label{eq.39c}
\sum_{k}p_{k}\vec{n}_{k}^{{}}=\sum_{k}\sqrt{p_{k}}\vec{N}_{k}.
\end{equation}
\begin{equation}\label{eq.40c}
\sum_{k}p_{k}\vec{n}_{k}^{\star }=\sum_{k}\sqrt{p_{k}}
\vec{N}_{k}^{\left( \star \right) }.\end{equation} The conditions
of solvability of the obtained equations is a criterion for
separability or entanglement of bipartite quantum spin state.

For Werner state \cite{JOBnew} with the density matrix
\begin{equation}\label{eq.41c}
\rho _{AB}=\left(
\begin{array}{cccc}
({1+p})/{4} & 0 & 0 & {p}/{2} \\ 0 & ({1-p})/{4} & 0 & 0 \\ 0 & 0
&({1-p})/{4} & 0 \\ {p}/{2} & 0 & 0 & ({1+p})/{4}
\end{array}
\right), \qquad \rho _{A}=\rho _{B}=\frac{1}{2}\left(
\begin{array}{cc}
1 & 0 \\
0 & 1
\end{array}
\right), \end{equation} one can reconstruct known results that for
$p<{1}/{3}$ the state is separable and for $p>{1}/{3}$ the state
is entangled, since in the decomposition of density operator in
the form (\ref{eq.26c}) the state
\begin{equation}\label{eq.42c}
\rho _{0}=\frac{1}{4}\left(
\begin{array}{cccc}
1 & 0 & 0 & 0 \\ 0 & 1 & 0 & 0 \\ 0 & 0 & 1 & 0 \\ 0 & 0 & 0 &
1\end{array} \right) \end{equation} has the weight
$p_{0}=(1-3p)/{4}$.

For $p>{1}/{3}$, the coefficient $p_{o}$ becomes negative.

There is some extension of the presented consideration.

Let us consider the state with the density matrix (nonnegative and
Hermitian)
\begin{equation}\label{W1}
\rho =\left(
\begin{array}{cccc}
R_{11} & 0 & 0 & R_{12} \\ 0 & \rho_{11} & \rho_{12} & 0 \\ 0 &
\rho_{21} & \rho_{22} & 0 \\ R_{21} & 0 & 0 & R_{22}\end{array}
\right), \qquad \mbox{Tr}\,\rho=1. \end{equation} Using procedure
of mapping the matrix onto vector $\vec\rho$ and applying to the
vector nonlocal linear transform corresponding to Peres partial
transpose and making inverse map of the transformed vector onto
the matrix, we obtain
\begin{equation}\label{W2}
\rho^m =\left(
\begin{array}{cccc}
R_{11} & 0 & 0 & \rho_{12} \\ 0 & \rho_{11} & R_{12} & 0 \\ 0 &
R_{21} & \rho_{22} & 0 \\ \rho_{21} & 0 & 0 & R_{22}\end{array}
\right). \end{equation} In the case of separable matrix $\rho$,
the matrix $\rho^m$ is nonnegative matrix. Calculating eigenvalues
of $\rho^m$ and applying condition of their positivity, we get
\begin{equation}\label{W3}
R_{11}R_{22}\geq |\rho_{12}|^2,\qquad  \rho_{11}\rho_{22}
\geq |R_{12}|^2.
\end{equation}
Violation of these inequalities gives a signal that $\rho$ is
entangled. For Werner state (\ref{eq.41c}), Eq.~(\ref{W3}) means
\begin{equation}\label{W4}
1+p>0,\qquad 1-p>2p,
\end{equation}
which recovers the condition of separability $p<1/3$ mentioned above.

The joint probability distribution (\ref{eq.25c}) of separable
state is positive after making the local and nonlocal (Peres-like)
transforms connected with positive map semigroup. But for
entangled state, function (\ref{eq.25c}) can take negative values
after making this map in the function. This is a criterion of
entanglement in terms of tomogram of the state of multiparticle
system.

\section{Dynamical Map and Purification}

In this section, we consider connection of positive maps with
purification procedure. In fact, formula
\begin{equation}\label{Pu1}
\rho\to\rho'=\sum_kp_kU_k\rho U_k^\dagger,\end{equation} where
$U_k$ are unitary operators, can be considered in the form
\begin{equation}\label{Pu2}
\rho\to\rho'=\sum_kp_k\rho_k,\qquad p_k\geq 0,\qquad \sum_kp_k=1.\end{equation}
Here the density operators $\rho_k$ read
\begin{equation}\label{Pu3}\rho_k=U_k\rho U_k^\dagger.\end{equation}
This form is the form of probabilistic addition. This mixture of
density operators can be purified
\begin{equation}\label{Pu4}\rho'\to\rho''=N\left[\sum_{kj}
\sqrt{p_kp_j}\,\frac{\rho_kP_0\rho_j}{\sqrt{
\mbox{Tr}\,\rho_kP_0\rho_jP_0}}\right],\end{equation}
where $P_0$ is a fiducial projector and
\begin{equation}\label{Pu5}
N^{-1}=\mbox{Tr}\,\left(\sum_{kj}\sqrt{p_kp_j}\,\frac{\rho_kP_0\rho_j}{\sqrt{
\mbox{Tr}\,\rho_kP_0\rho_jP_0}}\right].\end{equation} The map
(\ref{Pu1}) could be interpreted as the evolution in time of the
initial matrix $\rho_0$ considering unitary operators $U_k(t)$
depending on time. Thus one has
\begin{equation}\label{Pu6}
\rho_0\to\rho(t)=\sum_kp_kU_k(t)\rho_0U_k^\dagger(t).\end{equation}
In this case, the purification procedure provides the dynamical
map of a pure state
\begin{equation}\label{Pu7}
\mid\psi_0\rangle\langle\psi_0\mid\to\mid \psi(t)\rangle\langle
\psi(t)\mid,\end{equation} where $\mid \psi(t)\rangle$ obeys to a
nonlinear equation and, in the general case, this equation is not
differential equation in time variable like the Schr\"odinger
equation.

For a specific case, the evolution (\ref{Pu6}) can be described by
semigroup. The density matrix (\ref{Pu6}) obeys to first-order
differential equation in time for this case~[27--29].

The reason why there is no differential equation in time for generic case
is due to the absence of the property
\begin{equation}\label{Pu8}
\rho_{ij}(t_2)=\sum_{mn}K_{ij}^{mn}(t_2,t_1)\rho_{mn}(t_1),\end{equation}
where the kernel of evolution operator satisfies
\begin{equation}\label{Pu9}
K_{ij}^{mn}(t_3,t_2)K_{mn}^{pq}(t_2,t_1)=K_{ij}^{pg}(t_3,t_1).\end{equation}
It means  that trajectory (curve) is not determined by differential equation in time.

Thus via purification procedure and dynamical map of the density
matrix we get the dynamical map of a pure state (nonlinear
dynamical map). This map can be used in nonlinear models of
quantum motion.

\section{Density Matrix and Real Quadratic Forms}

It is convenient to associate the Hermitian nonnegative $n$$\times
$$n$ density matrix $\rho$ with real quadratic form determined by
the real matrix $D$ using the relationships
\begin{equation}\label{Q1}
r=\frac{\rho +\rho ^*}{2}, \qquad iR=\frac{\rho -\rho
^*}{2},\qquad r^{\rm t}=r,\qquad R^{\rm t}=-R,\end{equation} and
\begin{equation}\label{Q2}
D=\left(\begin{array}{cccc} r&R\\R^{\rm t}&r\end{array}
\right).\end{equation} The quadratic form is the scalar function
(homogeneous polynomial of second order)
\begin{equation}\label{Q3}
f\Big(\vec x,\vec y,r,R\Big)
=\Big(\vec x,\vec y\Big)D\left(
\begin{array}{c}
\vec x\\\vec y\end{array}\right)\geq 0.\end{equation}
Nonnegativity of $f\Big(\vec x,\vec y,r,R\Big)$ takes place for
all nonnegative Hermitian matrices $\rho$.

In fact, for complex vectors,
\begin{equation}\label{Q4}
\vec z=\vec x+i\vec y,\end{equation}
one has
\begin{equation}\label{Q5}
\vec z^*\rho\vec z=
f\Big(\vec x,\vec y,r,R\Big).\end{equation}
All real transforms of the form
\begin{equation}\label{Q6}
D\to D'=ADA^{\rm T}\end{equation}
keep the quadratic form nonnegative.

Let us consider the real $2n$$\times$$2n$ matrix $A$ given in block form
\begin{equation}\label{Q7}
A=\left(\begin{array}{cccc} a&b\\c&d\end{array}
\right),\end{equation} where $a,b,c$, and $d$ are real
$n$$\times$$n$ matrices.

If the matrix (\ref{Q6}) has the form
\begin{equation}\label{Q8}
D'=\left(\begin{array}{cccc}r'&R'\\ R'^{\rm t}&r'\end{array}
\right),\end{equation} one has the relationships:
\begin{eqnarray}
&&ara^{\rm t}+bR^{\rm t}a^{\rm t}+aRb^{\rm t}+brb^{\rm
t}=r',\label{Q9}\\ &&crc^{\rm t}+dR^{\rm t}c^{\rm t}+cRd^{\rm
t}+drd^{\rm t}=r'^{\rm t}, \label{Q10}\\ &&arc^{\rm t}+bR^{\rm
t}c^{\rm t}+aRd^{\rm t}+brd^{\rm t}=R',\label{Q11}\\ &&cra^{\rm
t}+dR^{\rm t}a^{\rm t}+cRb^{\rm t}+drb^{\rm t}=R'^{\rm t}.
\label{Q12}\end{eqnarray} Also $R'^{\rm t}=-R',\quad r'^{\rm
t}=r'$.

For the case
\begin{equation}\label{Q13}
b=c=0,
\end{equation}
one has possible solutions
\begin{equation}\label{Q14}
a=d\end{equation}
and
\begin{equation}\label{Q15}
a=-d.
\end{equation}
Solution (\ref{Q15}) for $a=1$ describes a Peres-like
transposition of the matrix $\rho$.

For the case $a=d=0$, one has possible solutions
\begin{equation}\label{Q16}
b=c\end{equation}
and
\begin{equation}\label{Q17}
b=-c.
\end{equation}
In the case (\ref{Q14}), one has
\begin{eqnarray}
&&r'=ara^{\rm t},\qquad R'=aRa^{\rm t}, \label{Q18}\\
&&r'+iR'=a(r+iR)a^{\rm t}.\label{Q19}
\end{eqnarray}
In the case (\ref{Q15}), one has
\begin{eqnarray}
&&r'=ara^{\rm t},\qquad R'=-aRa^{\rm t}, \label{Q20}\\
&&r'+iR'=a(r-iR)a^{\rm t}.\label{Q21}
\end{eqnarray}
Thus for block diagonal matrices $A$ the possible transforms of the initial
density matrix matrix $\rho$ have the form
\begin{equation}\label{Q22}
\rho\to\rho_a^{(\pm)}=a\left\{\begin{array}{c}
\rho\\\rho^{\rm t}\end{array}\right\}a^{\rm t}.
\end{equation}
In the vector form $\rho\leftrightarrow\vec\rho$,
the transform (\ref{Q22}) is described by superoperators
\begin{equation}\label{Q23}
L_a^{\pm}=\left\{\begin{array}{c} a\otimes a\\(a\otimes a)L^{\rm
t}\end{array}\right..\end{equation} Obviously one can apply the
averaging procedure to get the matrix $\langle a\otimes a\rangle$.
It is a partial case of the general transform $\langle v\otimes
v\rangle$ for real $v$. Here superpoperator $L^{\rm t}$ makes from
the vector $\vec\rho\leftrightarrow\rho$ the vector $\vec\rho_{\rm
t}\leftrightarrow\rho^{\rm t}$, where $\rho^{\rm t}$ is transposed
density matrix $\rho$.

The superoperator $L^{\rm t}$ in the case $n=2$ is described by
the matrix $g^{\alpha\beta}$. The solutions (\ref{Q16}) and
(\ref{Q17}) provide analogous transforms (\ref{Q22}) and
(\ref{Q23}) with replacement $a\to b$.

For $n=2$, the choice $a=\sigma_1$ gives
\begin{equation}\label{Q24}
\rho\to\sigma_1\rho^{\rm t}\sigma_1=\rho^{\rm t},
\end{equation}
which is exactly Peres transpose transform.

For $n=3$, the choice
\begin{equation}\label{Q25}
a=\left(\begin{array}{cccc}1&0&0\\0&-1&0\\0&0&-1\end{array}\right)\end{equation}
provides two transforms of the Hermitian density matrix
\begin{equation}\label{Q26}
\rho=\left(\begin{array}{cccc}\rho_{11}&\rho_{12}&\rho_{13}\\
\rho_{21}&\rho_{22}&\rho_{23}\\\rho_{31}&\rho_{32}&\rho_{33}
\end{array}\right)\to \rho'=\left(\begin{array}{cccc}\rho_{11}&\rho_{12}
&-\rho_{13}\\ \rho_{21}&\rho_{22}&-\rho_{23}\\-\rho_{31}&-\rho_{32}&\rho_{33}
\end{array}\right)
\end{equation}
and
\begin{equation}\label{Q27}
\rho \to \rho''=\left(\begin{array}{cccc}\rho_{11}&\rho_{21}
&-\rho_{31}\\ \rho_{12}&\rho_{22}&-\rho_{32}\\-\rho_{13}&-\rho_{23}&\rho_{33}
\end{array}\right).
\end{equation}
Obviously, the unitary transform $u$ of the form $$\rho\to u\rho
u^\dagger$$ does not change the nonnegative eigenvalues of the
density operator but this transform differs from the transforms
discussed above, e.g., if $|\mbox{det}\,a|\neq 1$, the above
transforms do not preserve the determinant of the density matrix.

The given construction can provide also noncompletely positive
map. For example, if the transform of Hermitian nonnegative
3$\times$3 matrix $\rho_{jk}$ is described by the formula
$$\rho_{jk}\to\rho_{jk}\Big(\cosh^2\theta\cos(\theta_j-\theta_k)
-\sinh^2\theta\delta_{jk}\Big)$$ (there is no sum over $j,k$), it
corresponds to applying to vector $\vec\rho$ the matrix $\langle
a\otimes a\rangle$ with real matrices $a$. The formula can be used
also for arbitrary integer $n$. But averaging is done using
quasidistribution (not the probability distribution). This means
that in sum $\sum_k\varepsilon_ka_k\otimes a_k$ the numbers
$\varepsilon_k$ take both positive and negative values $\pm 1$. It
is another example of noncompletely positive map. The example of
Peres transpose transform discussed in the previous sections for
the case $n=2$ belongs also to the case of positive but not
completely positive maps.

\section{Tomogram of the Group $U(n)$}

In order to formulate a criterion of separability for a bipartite
spin system with spin $j_1$ and $j_2$, we introduce the tomogram
$w(\vec l, \vec m,g^{(n)})$ for the group $U(n)$, where
$$n=n_1n_2,\qquad n_1=2j_1+1,\qquad n_2=2j_2+1,$$ and $g^{(n)}$
are parameters of the group element. Vectors $\vec l$ and $\vec m$
label a basis $\mid \vec l,\vec m\rangle$ of the fundamental
representation of the group $U(n)$. For example, since this
representation is irreducible, being reduced to representation of
$U(n_1)\otimes U(n_2)$-subgroup of the group $U(n)$, the basis can
be chosen as the product of basis vectors:
\begin{equation}\label{U1}
\mid j_1,m_1\rangle \mid j_2,m_2\rangle =\mid
j_1,j_2,m_1,m_2\rangle.
\end{equation}
Due to irreducibility of this representation of the group $U(n)$
and its subgroup, there exists a unitary transform
$u_{j_1j_2m_1m_2}^{\vec l\vec m}\mid \vec l,\vec m\rangle$ such
that
\begin{eqnarray}
&&\mid j_1,j_2,m_1,m_2\rangle=\sum_{\vec l\vec m}
u_{j_1j_2m_1m_2}^{\vec l\vec m}\mid \vec l,\vec
m\rangle,\label{U2}\\ &&\mid \vec l\vec m
\rangle=\sum_{m_1m_2}(u^{-1})^{\vec l\vec m}_{j_1j_2m_1m_2}\mid
j_l,j_2,m_1,m_2\rangle. \label{U3}\end{eqnarray} One can define
the $U(n)$-tomogram for a Hermitian nonnegative $n$$\times$$n$
density matrix $\rho$, which belongs to Lie algebra of the group
$U(n)$, by a generic formula:
\begin{equation}\label{U4}
w(\vec l,\vec m,g^{(n)})=\langle\vec l,\vec m\mid
U^\dagger(g^{(n)})\rho U(g^{(n)})\mid\vec l,\vec m\rangle.
\end{equation}
Formula~(\ref{U4}) defines the tomogram in basis $\mid\vec l,\vec
m\rangle$.

Now let us define the $U(n)$-tomogram using basis $\mid
j_1,j_2,m_1,m_2\rangle$, i.e.,
\begin{equation}\label{U5}
w^{(j_1,j_2)}(m_1, m_2,g^{(n)})=\langle j_1,j_2,m_1,m_2\mid
U^\dagger(g^{(n)})\rho U(g^{(n)})\mid j_1,j_2,m_1,m_2\rangle.
\end{equation}
This tomogram is spin-tomogram~\cite{Andreev} for $g^{(n)}\in
U(2)\otimes U(2)$ subgroup of the group $U(n)$. Properties of this
tomogram follow from its meaning to be joint probability
distribution of two random spin projections $m_1,m_2$ depending on
$g^{(n)}$ parameters.

One has normalization condition
\begin{equation}\label{U6}
\sum_{m_1,m_2}w^{(j_1,j_2)}(m_1, m_2,g^{(n)})=1.
\end{equation}
Also all the probabilities are nonnegative, i.e.,
\begin{equation}\label{U7}
w^{(j_1,j_2)}(m_l, m_2,g^{(n)})\geq 0.
\end{equation}
Due to this, one has
\begin{equation}\label{U8}
\sum_{m_1,m_2}|w^{(j_1,j_2)}(m_l, m_2,g^{(n)})|=1.
\end{equation}
For spin-tomogram,
\begin{equation}\label{U9}
g^{(n)}\rightarrow \Big(\vec O_1,\vec O_2\Big)
\end{equation}
and
\begin{equation}\label{U10}
w^{(j_1,j_2)}(m_l, m_2,g^{(n)})\rightarrow w(m_1,m_2,\vec O_1,\vec
O_2).
\end{equation}

The separability and entanglement condition discussed in the
previous section for bipartite spin-tomogram can be considered
also from the viewpoint of the properties of $U(n)$-tomogram. If
the two-spin $n$$\times$$n$ density matrix $\rho$ is separable, it
keeps to be separable under action of generic positive map of the
subsystem density matrices. This map can be described as follows.

Let $\rho$ to be mapped onto vector $\vec \rho$ with $n^2$
components. The components are simply ordered rows of the matrix
$\rho$, i.e.,
\begin{equation}\label{U11}
\vec\rho=\Big(\rho_{11},\rho_{12},\ldots,\rho_{1n},\rho_{21},\rho_{22},
\ldots,\rho_{nn},\Big).
\end{equation}
The $n^2$$\times$$n^2$ matrix $L$ is taken in the form
\begin{equation}\label{U12} L=\sum_sp_sL_s^{(j_1)}\otimes
L_s^{(j_2)},\qquad p_s\geq 0,\quad \sum_s p_s=1,
\end{equation}
where $n_1$$\times$$n_1$ matrix $L_s^{(j_1)}$ and
$n_2$$\times$$n_2$ matrix $L_s^{(j_2)}$ describe the positive maps
of density matrices of spin-$j_1$ and spin-$j_2$ subsystems,
respectively. We map vector $\vec\rho$ onto vector $\vec\rho_L$
\begin{equation}\label{U13}
\vec\rho_L=L\vec\rho
\end{equation}
and construct the $n^2$$\times$$n^2$ matrix $\rho_L$, which
corresponds to the vector $\vec\rho_L$. Then we consider
$U(n)$-tomogram of the matrix $\rho_L$, i.e.,
\begin{equation}\label{U14}
w^{(j_1,j_2)}_L(m_l, m_2,g^{(n)})=\langle j_1,j_2,m_1,m_2\mid
U^\dagger(g^{(n)}\rho_L U(g^{(n)})\mid j_1,j_2,m_l,m_2\rangle.
\end{equation}
Using this tomogram we introduce the function
\begin{equation}\label{U15}
F(g^{(n)},L)=\sum_{m_1,m_2}\left|w_L^{(j_1,j_2)}(m_1,m_2,g^{(n)})\right|.
\end{equation} For separable states, this function does not depend
on the $U(n)$-group parameter $g^{(n)}$ and positive-map matrix
elements of the matrix $L$.

For normalized density matrix $\rho$ of the bipartite spin-system,
this function reads
\begin{equation}\label{U16}
F(g^{(n)},L)=1.
\end{equation}
For entangled states, this function depends on $g^{(n)}$ and $L$
and it is not equal to unity. This property can be chosen as
necessary and sufficient condition for separability of bipartite
spin-states. In fact, the formulated approach can be extended to
multipartite systems too. The generalization is as follows.

Given $N$ spin-systems with spins $j_1,j_2,\ldots,j_N$. Let us
consider the group $U(n)$ with
\begin{equation}\label{U17}
n=\prod_{k=1}^Nn_k,\qquad n_k=2j_k+1.
\end{equation}
Let us introduce basis
\begin{equation}\label{U18}
\mid \vec m\rangle=\prod_{k=1}^N\mid j_km_k\rangle
\end{equation}
in the linear space of the fundamental representation of the group
$U(n)$. We define now $U(n)$-tomogram of a state with
$n^2$$\times$$n^2$ matrix $\rho$:
\begin{equation}\label{U19}
w_\rho(\vec m,g^{(n)})=\langle \vec m\mid U^\dagger(g^{(n)})\rho
U(g^{(n)})\mid\vec m\rangle.
\end{equation}
For positive Hermitian matrix $\rho$ with $\mbox{Tr}\,\rho=1$, we
formulate a criterion of separability as follows.

Let the map matrix $L$ to be of the form
\begin{equation}\label{U20}
L=\sum_sp_s\Big(\prod_{k=1}^N\otimes L_s^{(k)}\Big), \qquad
p_s\geq 0,\quad \sum_s p_s=1,
\end{equation}
where $L_s^{(k)}$ is positive-map matrix of the density matrix of
$k$th spin subsystem. We construct the matrix $\rho_L$ as in the
case of bipartite system using the matrix $L$. The function
\begin{equation}\label{U21}
F(g^{(n)},L)=\sum_{\vec m}|w_{\rho_L}(\vec m,g^{(n)})|\geq 1
\end{equation}
is equal to unity for separable state and it depends on the matrix
$L$ and $U(n)$-parameters $g^{(n)}$ for entangled states. This
criterion can be applied also in the case of continuous variables,
e.g., for Gaussian states of photons. Function (\ref{U21}) can
provide the measure of entanglement. Thus one can use maximum
value (or a mean value) of this function as a characteristic of
entanglement. In fact, the separability criterion is related to
the following positivity criterion of finite or infinite matrix
$A$. The matrix $A$ is positive iff the sum of moduli of diagonal
matrix elements of the matrix $UAU^\dagger$ is equal to positive
trace of the matrix $A$ for arbitrary unitary matrix $U$.

\section{Conclusions}

To conclude, we formulated the notion of separability and
entanglement as a criterion for joint tomographic probability of
subsystem states to be represented in the specific form of sum of
products of tomograms of the subsystems.

We have shown that the positive map of density matrix of
multiparticle system expressed in terms of superoperator acting in
Lie algebra (adjoint representation) of unitary group can be
considered as a semigroup, which contains all local unitary
transforms acting in subspaces corresponding to the subsystem
states.

The set of separable states is shown to be invariant under action of this group.

The intrinsic measure of entanglement is shown to be invariant
under action of the local group.

The formalism of vectors representing the matrices is convenient
tool for the consideration. We introduced unitary spin tomogram
and formulated necessary and sufficient condition of entanglement.

\section*{Acknowledgments}

V~I~M and E~C~G~S thank Dipartimento di Scienze Fisiche,
Universit\'a ``Federico~II'' di Napoli and Istitito Nazionale di
Fisica Nucleare, Sezione di Napoli for kind hospitality. V~I~M is
grateful to the Russian Foundation for Basic Research for partial
support under Project~No.~01-02-17745.

\end{document}